\documentclass[reprint,superscriptaddress,preprintnumbers,nofootinbib,amsmath,amssymb,aps,prd,showkeys,showpacs]{revtex4-2}

\usepackage{graphicx}
\usepackage{dcolumn}
\usepackage{bm}
\usepackage{array} 
\newcolumntype{M}[1]{>{\centering\arraybackslash}m{#1}}
\usepackage{endnotes} 
\usepackage{adjustbox}

\usepackage{amssymb}    
\usepackage{color}         
\usepackage{graphicx}     
\usepackage{mathrsfs} 
\usepackage{hyperref} 
\hypersetup{colorlinks=true,linkcolor=blue,citecolor=blue}
\usepackage{ upgreek } 
\usepackage{color,xcolor,fancybox,epsf,rotating,colordvi}
\usepackage{booktabs}
\usepackage{multirow}

\usepackage{soul}
\setstcolor{red}

\usepackage{cancel}

\begin{document}

\title{Probing dark matter below the neutrino floor through NMSSM Higgs cascade decays at the HL-LHC}

\author{Yabo Dong}
\affiliation{School of Physics and Electronics, Henan University, Kaifeng 475004, China}

\author{Kun Wang}
\email[Corresponding author:]{kwang@usst.edu.cn} 
\affiliation{College of Science, University of Shanghai for Science and Technology, Shanghai 200093, China}

\author{Haijun Yang}
\affiliation{State Key Laboratory of Dark Matter Physics, Key Laboratory for Particle Astrophysics and Cosmology (MOE), Shanghai Key Laboratory for Particle Physics and Cosmology (SKLPPC),  School of Physics and Astronomy \mbox{\normalfont\&} Tsung-Dao Lee Institute, Shanghai Jiao Tong University, Shanghai 200240}

\author{Jingya Zhu}
\email[Corresponding author:]{zhujy@henu.edu.cn} 
\affiliation{School of Physics and Electronics, Henan University, Kaifeng 475004, China}


\date{\today}

\begin{abstract}
We investigate the potential of Higgs cascade decays to probe neutralino dark matter below the neutrino floor in the semi-constrained Next-to-Minimal Supersymmetric Standard Model with nonuniversal gaugino masses. We focus on lightest-neutralino masses of 65--100~GeV and study the process $pp\to H_2\to A_1A_1\to\gamma\gamma+\tilde{\chi}_1^0\tilde{\chi}_1^0$, where $H_1$ is the observed SM-like Higgs boson, $H_2$ is a heavier CP-even state, and $A_1$ is the lightest CP-odd state. 
The selected parameter region lies near the $A_1$ resonance, with $m_{A_1}>2m_{\tilde{\chi}_1^0}$, allowing both resonantly enhanced neutralino annihilation and invisible $A_1$ decays. 
After imposing theoretical and experimental constraints, we find surviving samples with cascade signal cross sections of approximately 0.16~fb at $\sqrt{s}=14~\mathrm{TeV}$ even when their relic-density-rescaled spin-independent scattering cross sections lie below the neutrino floor. 
A Monte Carlo analysis of a representative benchmark point yields an expected statistical significance of approximately $6.1\sigma$ at the high-luminosity LHC with an integrated luminosity of $3000~\mathrm{fb}^{-1}$, neglecting systematic uncertainties.
These results highlight the potential of the diphoton plus missing transverse momentum channel as a complementary probe of neutralino dark matter.
\end{abstract}

\maketitle
\newpage



\section{Introduction}
\label{sec:intro}

Recent results from the LUX-ZEPLIN (LZ) experiment have placed stringent limits on the spin-independent scattering cross section of weakly interacting massive particles (WIMPs) with nucleons. With an exposure of 4.2~tonne$\cdot$years, LZ reported an upper limit of $2.2\times10^{-48}~\mathrm{cm}^2$ at a WIMP mass of 40~GeV at 90\% confidence level~\cite{LZ:2024zvo}. As direct detection experiments improve their sensitivity, coherent elastic neutrino-nucleus scattering becomes an increasingly important background, making further gains in sensitivity more challenging in the region commonly referred to as the neutrino floor~\cite{Billard:2013qya}. This motivates complementary searches at colliders, where invisible particles can be produced and probed through signatures involving missing transverse momentum~\cite{Arcadi:2017kky,Boveia:2018yeb}. In particular, the high-luminosity LHC (HL-LHC) offers an opportunity to explore dark matter scenarios with scattering cross sections below the neutrino floor.

A widely used strategy at the LHC is to search for mono-$X$ signatures, in which a visible Standard Model (SM) object, such as a jet, a photon, or a $W/Z$ boson, recoils against invisible particles. Monojet and monophoton searches have placed stringent constraints on dark matter production in a variety of simplified models~\cite{ATLAS:2021kxv,ATLAS:2020uiq}. However, their interpretation in terms of dark matter--nucleon scattering depends on the mediator properties and the assumed couplings~\cite{Abdallah:2015ter}. Higgs-portal scenarios provide a connection between dark matter phenomenology and an extended scalar sector~\cite{Baum:2017enm}. ATLAS has searched for dark matter produced in association with the 125~GeV Higgs boson in the diphoton plus missing transverse momentum final state~\cite{ATLAS:2021jbf}. Meanwhile, invisible decays of the observed Higgs boson into a dark matter pair are kinematically forbidden when the dark matter mass exceeds half its mass. These considerations motivate complementary searches involving additional Higgs bosons, whose cascade decays can produce visible resonances together with missing transverse momentum.

Supersymmetric extensions of the SM provide a well-motivated framework for studying such signatures. If $R$ parity is conserved and the lightest neutralino $\tilde{\chi}_1^0$ is the lightest supersymmetric particle (LSP), it is stable and can constitute part or all of the dark matter~\cite{Jungman:1995df}. In the Minimal Supersymmetric Standard Model (MSSM), the neutral Higgs sector contains two CP-even states, $h$ and $H$, and one CP-odd state, $A$, which can mediate neutralino pair production~\cite{Djouadi:2005gj}. An SM-like Higgs boson $h$ with a mass of approximately 125~GeV can decay invisibly into two on-shell neutralinos only for $m_{\tilde{\chi}_1^0}<m_h/2$. The prospects for probing neutralino dark matter through this decay have been studied~\cite{Cao:2012im}, and its branching fraction is constrained by LHC searches for invisible Higgs decays~\cite{ATLAS:2023tkt}. For heavier neutralinos, additional Higgs bosons provide alternative production channels. One possible topology is $pp\to H\to AA\to\gamma\gamma+\tilde{\chi}_1^0\tilde{\chi}_1^0$, provided that the required mass hierarchy is realized. The interactions relevant to this topology are already present in the MSSM, although its viability and observable rate depend on the Higgs mass spectrum, the neutralino composition, and the competing Higgs decay modes. This motivates investigating the diphoton plus missing transverse momentum signature in supersymmetric models with an extended Higgs sector.

Naturalness considerations motivate supersymmetric scenarios with an electroweak sector near the weak scale~\cite{Hall:2011aa,Baer:2012uy,Cao:2018rix}. Although direct searches have placed stringent constraints on colored superpartners~\cite{ATLAS:2020syg,CMS:2019zmd}, additional Higgs bosons with masses of a few hundred GeV remain possible in suitable regions of parameter space. The Next-to-Minimal Supersymmetric Standard Model (NMSSM) extends the MSSM with a gauge-singlet chiral superfield, providing additional freedom in the Higgs and neutralino sectors~\cite{Ellwanger:2009dp,Maniatis:2009re,Miller:2003ay,Li:2022etb,Du:2017str,Du:2018pko,Lian:2024smg,Carena:2015moc,Zhou:2021pit,Ellwanger:2024txc,Cao:2023gkc}. Its Higgs sector can accommodate an SM-like state near 125~GeV together with additional neutral Higgs bosons~\cite{King:2012is,Cao:2012fz}. Higgs cascade decays have been investigated as a means of searching for these additional states~\cite{King:2014xwa,Ellwanger:2017skc,Baum:2019uzg,Ellwanger:2022jtd}, and their potential to yield mono-Higgs signatures has been studied in the NMSSM~\cite{Baum:2017gbj}.

Decays of a Higgs boson into pairs of lighter singlet-dominated scalars or pseudoscalars have been explored in the NMSSM and scNMSSM~\cite{Dermisek:2005ar,Cao:2013gba,Ma:2020mjz}. In the configuration of interest here, a singlet-dominated CP-odd Higgs boson can be lighter than the additional CP-even Higgs states, allowing decays of the latter into a pair of pseudoscalars when kinematically accessible. Its suppressed couplings to SM fermions can reduce competing fermionic decay widths, while chargino loops can generate a diphoton decay amplitude. An appreciable diphoton branching fraction can therefore be obtained for suitable Higgs and chargino compositions~\cite{Guchait:2016pes}. If the pseudoscalar mass lies above twice the lightest neutralino mass, its invisible decay into a neutralino pair is also allowed, with a branching fraction determined by the neutralino composition, the available phase space, and the competing decay modes. Near the resonance condition $m_A\approx2m_{\tilde{\chi}_1^0}$, pseudoscalar-mediated annihilation can play an important role in determining the neutralino relic density~\cite{Gunion:2005rw}. More generally, the interplay between light neutralino dark matter and the Higgs sector has been studied in the NMSSM~\cite{Cao:2011re,Li:2023kbf} and the general NMSSM (GNMSSM)~\cite{Li:2025qkg}. These connections motivate studying Higgs cascade decays into a diphoton pair and missing transverse momentum together with the dark matter constraints.

Nonuniversal gaugino masses at the grand unification (GUT) scale allow a heavy gluino to coexist with relatively light electroweak gauginos, as illustrated in extensions of the CMSSM~\cite{Wang:2018vrr,Dong:2024juh}.
In this work, we study the semi-constrained NMSSM (scNMSSM) with nonuniversal gaugino masses, in which $M_1$, $M_2$, and $M_3$ are independent inputs at the GUT scale. The Higgs sector contains three CP-even states, $H_{1,2,3}$, two CP-odd states, $A_{1,2}$, and a pair of charged states, $H^\pm$, with the neutral states ordered by increasing mass within each CP sector. We focus on scenarios in which $H_1$ is the observed SM-like Higgs boson and the lightest neutralino has a mass between 65 and 100~GeV. 
We consider surviving samples satisfying $m_{A_1}>2m_{\tilde{\chi}_1^0}$, for which the decay $A_1\to\tilde{\chi}_1^0\tilde{\chi}_1^0$ is kinematically allowed. 
The cascade $gg\to H_2\to A_1A_1\to\gamma\gamma+\tilde{\chi}_1^0\tilde{\chi}_1^0$ becomes kinematically allowed when $m_{H_2}>2m_{A_1}$. After imposing theoretical and experimental constraints, we examine the neutralino composition, the dominant dark matter annihilation channels, and the signal production rate. We then perform Monte Carlo simulations of the signal and relevant SM backgrounds for a representative benchmark point at the 14~TeV HL-LHC to estimate the sensitivity of the diphoton plus missing transverse momentum channel. Our analysis explores the potential of this channel to probe scenarios in which the relic-density-rescaled spin-independent neutralino--nucleon scattering cross section lies below the neutrino floor.

The remainder of this paper is organized as follows. 
Section~\ref{sec:model} introduces the scNMSSM framework and the theoretical and experimental constraints used in our analysis. 
Section~\ref{sec:scan} describes the parameter scan and discusses the properties of the surviving samples, including their dark matter phenomenology and signal production rates. 
Section~\ref{sec:MC} presents the Monte Carlo simulations of the signal and SM backgrounds and estimates the sensitivity at the HL-LHC. 
Finally, our conclusions are summarized in Sec.~\ref{sec:conclusion}.

\section{Model framework and constraints}
\label{sec:model}
The NMSSM extends the MSSM by a gauge-singlet chiral superfield $\hat{S}$. The $Z_3$-invariant superpotential is given by~\cite{Ellwanger:2009dp}
\begin{equation}
\begin{aligned}
W_{\mathrm{NMSSM}} ={}& y_u \hat{Q}\cdot\hat{H}_u \hat{u}^c + y_d \hat{H}_d\cdot\hat{Q} \hat{d}^c + y_e \hat{H}_d\cdot\hat{L} \hat{e}^c \\
&+ \lambda \hat{S}\hat{H}_u\cdot\hat{H}_d + \frac{\kappa}{3}\hat{S}^3,
\end{aligned}
\end{equation}
where $\lambda$ and $\kappa$ are dimensionless couplings, the dot denotes the antisymmetric $SU(2)_L$ contraction, and generation indices are suppressed. When the scalar component of $\hat{S}$ acquires a vacuum expectation value (VEV), $v_s\equiv\langle S\rangle$, an effective Higgsino mass parameter is generated,
\begin{equation}
\mu_{\mathrm{eff}}=\lambda v_s.
\end{equation}
The VEVs of the neutral doublet Higgs fields are denoted by $v_u\equiv\langle H_u^0\rangle$ and $v_d\equiv\langle H_d^0\rangle$, with
\begin{equation}
\tan\beta=\frac{v_u}{v_d},
\end{equation}
where $v_u^2+v_d^2=v^2\approx(174~\mathrm{GeV})^2$.

The soft SUSY-breaking terms of the $Z_3$-invariant NMSSM are given by~\cite{Ellwanger:2009dp}
\begin{equation}
\begin{aligned}
-\mathcal{L}_{\mathrm{NMSSM}}^{\mathrm{soft}} ={}& -\left.\mathcal{L}_{\mathrm{MSSM}}^{\mathrm{soft}}\right|_{B\mu=0} + m_S^2 |S|^2 \\
&+ \left(\lambda A_\lambda S H_u\cdot H_d + \frac{\kappa}{3}A_\kappa S^3 + \mathrm{h.c.}\right),
\end{aligned}
\end{equation}
where $\left.\mathcal{L}_{\mathrm{MSSM}}^{\mathrm{soft}}\right|_{B\mu=0}$ denotes the MSSM soft SUSY-breaking Lagrangian with the Higgs bilinear soft term omitted. The fields $S$, $H_u$, and $H_d$ are the scalar components of the corresponding chiral superfields, $m_S^2$ is the singlet soft mass-squared parameter, and $A_\lambda$ and $A_\kappa$ are dimensionful trilinear soft parameters.

In this work, we consider a semi-constrained NMSSM with nonuniversal gaugino masses~\cite{Li:2022etb}. 
At the GUT scale, a common soft mass $M_0$ is assigned to the sfermions, and their trilinear soft parameters are unified to $A_0$. 
The gaugino masses $M_1$, $M_2$, and $M_3$ are treated as independent inputs, while the Higgs-sector trilinear parameters $A_\lambda$ and $A_\kappa$ are allowed to differ from $A_0$. 
The model is parametrized by the following 11 inputs:
\begin{equation}
\begin{gathered}
M_0,\ M_1,\ M_2,\ M_3,\ A_0,\ A_\lambda,\ A_\kappa,\\
\tan\beta,\ \lambda,\ \kappa,\ \mu_{\mathrm{eff}}.
\end{gathered}
\end{equation}
Here $A_\lambda$ and $A_\kappa$ are specified at the GUT scale, while $\lambda$, $\kappa$, and $\mu_{\mathrm{eff}}$ are defined at the SUSY scale $M_{\mathrm{SUSY}}$ used by NMSSMTools, and $\tan\beta$ is defined at $M_Z$.

After electroweak symmetry breaking, the scalar Higgs fields can be expanded around their VEVs as
\begin{equation}
\begin{aligned}
H_u &= \begin{pmatrix} H_u^+ \\ v_u + \dfrac{H_u^R+iH_u^I}{\sqrt{2}} \end{pmatrix},\\
H_d &= \begin{pmatrix} v_d + \dfrac{H_d^R+iH_d^I}{\sqrt{2}} \\ H_d^- \end{pmatrix},\\
S &= v_s + \frac{S^R+iS^I}{\sqrt{2}},
\end{aligned} 
\end{equation}
where $H_u^R$, $H_d^R$, and $S^R$ denote the CP-even neutral field fluctuations, while $H_u^I$, $H_d^I$, and $S^I$ denote the CP-odd ones in the CP-conserving limit. 
The fields $H_u^+$ and $H_d^-$ are the charged components of the Higgs doublets.

It is convenient to rotate the two Higgs doublets to the Higgs basis~\cite{Cao:2016uwt},
\begin{align}
\mathcal{H}_1
&= \cos\beta\,H_u+\sin\beta\,\varepsilon H_d^*
= \begin{pmatrix}
H^+ \\
\dfrac{S_1+iP_1}{\sqrt{2}}
\end{pmatrix},
\notag\\[4pt]
\mathcal{H}_2
&= \sin\beta\,H_u-\cos\beta\,\varepsilon H_d^*
= \begin{pmatrix}
G^+ \\
v+\dfrac{S_2+iG^0}{\sqrt{2}}
\end{pmatrix},
\end{align}
where $\varepsilon=i\sigma_2=\begin{pmatrix}0&1\\-1&0\end{pmatrix}$ is the antisymmetric $SU(2)$ tensor. 
Only $\mathcal{H}_2$ acquires a VEV, and $S_2$ is the neutral CP-even fluctuation along the electroweak symmetry-breaking direction. 
The fields $G^\pm$ and $G^0$ are the would-be Goldstone bosons, while $H^\pm$ denotes the physical charged Higgs boson. The singlet components are denoted by $S_3\equiv S^R$ and $P_2\equiv S^I$. 
The doublets $\mathcal{H}_{1,2}$ should be distinguished from the neutral CP-even mass eigenstates $H_{1,2,3}$.

In the CP-conserving limit, the neutral Higgs mass eigenstates are obtained by diagonalizing the CP-even and CP-odd squared-mass matrices, $M_S^2$ and $M_P^2$, in the bases $(S_1,S_2,S_3)$ and $(P_1,P_2)$, respectively. 
Explicit tree-level expressions for $M_S^2$ and $M_P^2$ in these bases can be found in Refs.~\cite{Cao:2016uwt} and~\cite{Ellwanger:2009dp}, respectively. 
The mass eigenstates are related to the basis fields by
\begin{equation}
\begin{aligned}
H_i &= \sum_{j=1}^{3}\mathcal{S}_{ij}S_j,\qquad i=1,2,3,\\
A_i &= \sum_{j=1}^{2}\mathcal{P}_{ij}P_j,\qquad i=1,2,
\end{aligned}   
\end{equation}
where the real orthogonal matrices $\mathcal{S}$ and $\mathcal{P}$ satisfy
\begin{equation}
\begin{aligned}
\mathcal{S}M_S^2\mathcal{S}^{T} &= \operatorname{diag}(m_{H_1}^2,m_{H_2}^2,m_{H_3}^2),\\
\mathcal{P}M_P^2\mathcal{P}^{T} &= \operatorname{diag}(m_{A_1}^2,m_{A_2}^2).
\end{aligned}    
\end{equation}
The states are ordered by increasing mass, $m_{H_1}\leq m_{H_2}\leq m_{H_3}$ and $m_{A_1}\leq m_{A_2}$. In this basis, $|\mathcal{S}_{i3}|^2$ and $|\mathcal{P}_{i2}|^2$ give the singlet fractions of $H_i$ and $A_i$, respectively.

The neutralino sector contains the bino, the neutral wino, the two neutral higgsinos, and the singlino. In the gauge-eigenstate basis $\psi^0=(-i\tilde{B},-i\tilde{W}^0,\tilde{H}_d^0,\tilde{H}_u^0,\tilde{S})^T$, the neutralino mass term is written as~\cite{Ellwanger:2009dp}
\begin{equation}
\mathcal{L}_{\mathrm{mass}}=-\frac{1}{2}(\psi^0)^T M_{\tilde{\chi}^0}\psi^0+\mathrm{h.c.},
\end{equation}
with
\begin{widetext}
\begin{equation}
\label{eq:neutralino_mass}
M_{\tilde{\chi}^0}=
\begin{pmatrix}
M_1 & 0 & -c_\beta s_W m_Z & s_\beta s_W m_Z & 0 \\
0 & M_2 & c_\beta c_W m_Z & -s_\beta c_W m_Z & 0 \\
-c_\beta s_W m_Z & c_\beta c_W m_Z & 0 & -\mu_{\mathrm{eff}} & -\lambda v_u \\
s_\beta s_W m_Z & -s_\beta c_W m_Z & -\mu_{\mathrm{eff}} & 0 & -\lambda v_d \\
0 & 0 & -\lambda v_u & -\lambda v_d & 2\kappa v_s
\end{pmatrix}.
\end{equation}
\end{widetext}
Here $s_\beta=\sin\beta$, $c_\beta=\cos\beta$, $s_W=\sin\theta_W$, and $c_W=\cos\theta_W$, where $\theta_W$ is the weak mixing angle. 
The parameters in this matrix are evaluated at the low-energy renormalization scale; in particular, $M_1$ and $M_2$ are the running gaugino masses obtained from the GUT-scale inputs. 
The neutralino mass eigenstates are defined by $\tilde{\chi}_i^0=N_{ij}\psi_j^0$, where $N$ is a unitary matrix satisfying
\begin{equation}
N^*M_{\tilde{\chi}^0}N^\dagger=\operatorname{diag}(m_{\tilde{\chi}_1^0},\ldots,m_{\tilde{\chi}_5^0}),
\end{equation}
with nonnegative masses ordered increasingly. 
The bino, wino, higgsino, and singlino fractions of the lightest neutralino are given by $|N_{11}|^2$, $|N_{12}|^2$, $|N_{13}|^2+|N_{14}|^2$, and $|N_{15}|^2$, respectively.

We impose the following theoretical and experimental constraints using NMSSMTools~6.2.0~\cite{Ellwanger:2004xm,Ellwanger:2005dv,Ellwanger:2006rn,Das:2011dg,Muhlleitner:2003vg}, supplemented by HiggsTools and the subsequent sparticle-search checks described below.

\begin{itemize}

\item \textit{Theoretical constraints.} We require successful electroweak symmetry breaking, the absence of tachyonic states and Landau poles below the GUT scale, and a neutralino LSP. The vacuum constraints implemented in NMSSMTools are also imposed.

\item \textit{Higgs-sector constraints.} We apply the Higgs constraints implemented in NMSSMTools and further test the samples using HiggsBounds~\cite{Bechtle:2008jh} and HiggsSignals~\cite{Bechtle:2013xfa} within HiggsTools~\cite{Bahl:2022igd}. The HiggsSignals analysis includes 159 observables, and we require $\chi^2_{\mathrm{HS}}<190$. For the cascade analysis, $H_1$ is identified as the observed SM-like Higgs boson, with $124~\mathrm{GeV}<m_{H_1}<126~\mathrm{GeV}$.

\item \textit{Sparticle searches.} We apply the LEP, Tevatron, and internal CMS electroweakino constraints implemented in NMSSMTools~\cite{CMS:2018szt}. Squark and gluino masses are required to be at least 1~TeV during the scan. The surviving samples are subsequently checked using SModelS~\cite{Kraml:2013mwa,Altakach:2024jwk} and the additional slepton constraints implemented in NMSSMTools~\cite{ATLAS:2019lng,ATLAS:2022hbt}.

\item \textit{Flavor constraints.} We impose the constraints from $\Upsilon$ decays and $B$- and $K$-meson observables implemented in NMSSMTools~\cite{Domingo:2007dx,Domingo:2015wyn,Domingo:2008rr,Domingo:2010am}. The experimental input intervals for three representative branching fractions are
\begin{equation}
\begin{aligned}
&\mathrm{BR}(B\to X_s\gamma)_{\mathrm{exp}} \in [3.02,3.62]\times10^{-4},\\
&\mathrm{BR}(B_s\to\mu^+\mu^-)_{\mathrm{exp}} \in [2.1,3.7]\times10^{-9},\\
&\mathrm{BR}(B^+\to\tau^+\nu_\tau)_{\mathrm{exp}} \in [0.78,1.44]\times10^{-4}.
\end{aligned}  
\end{equation}

\item \textit{Dark matter relic density.} The neutralino relic density is calculated using micrOMEGAs integrated into NMSSMTools~\cite{Belanger:2005kh,Belanger:2008sj,Alguero:2023zol}. We impose the upper bound $\Omega_{\tilde{\chi}_1^0}h^2<0.131$, allowing the neutralino to constitute part or all of the dark matter~\cite{Planck:2015fie}.

\item \textit{Dark matter direct detection.} We apply the spin-independent and spin-dependent limits implemented in NMSSMTools, including LZ~\cite{LZ:2024zvo,LZ:2025igz}, PandaX-4T~\cite{PandaX:2025rrz}, XENONnT~\cite{XENONCollaborationP:2026ioh}, and DarkSide-50~\cite{DarkSide-50:2025lns}. For a subdominant neutralino component, the local density is assumed to scale with its cosmological fraction. We present the spin-independent scattering results as $\xi\sigma_{\mathrm{SI}}$, where $\xi=\Omega_{\tilde{\chi}_1^0}h^2/0.1188$\footnote{LZ has also investigated covariant WIMP--nucleon effective interactions~\cite{LZ:2024vge} and, more recently, effective-field-theory and inelastic dark matter scenarios in an extended nuclear-recoil energy window~\cite{LZ:2026axp}. Our analysis uses the standard elastic spin-independent and spin-dependent neutralino--nucleon scattering limits. Bounds on additional effective operators and inelastic transitions require a model-specific reinterpretation and are not imposed as separate constraints on our samples.}.

\end{itemize}

\section{Parameter scan and phenomenology}
\label{sec:scan}

\begin{figure*}[!ht]
    \centering
    \includegraphics[width=\linewidth]{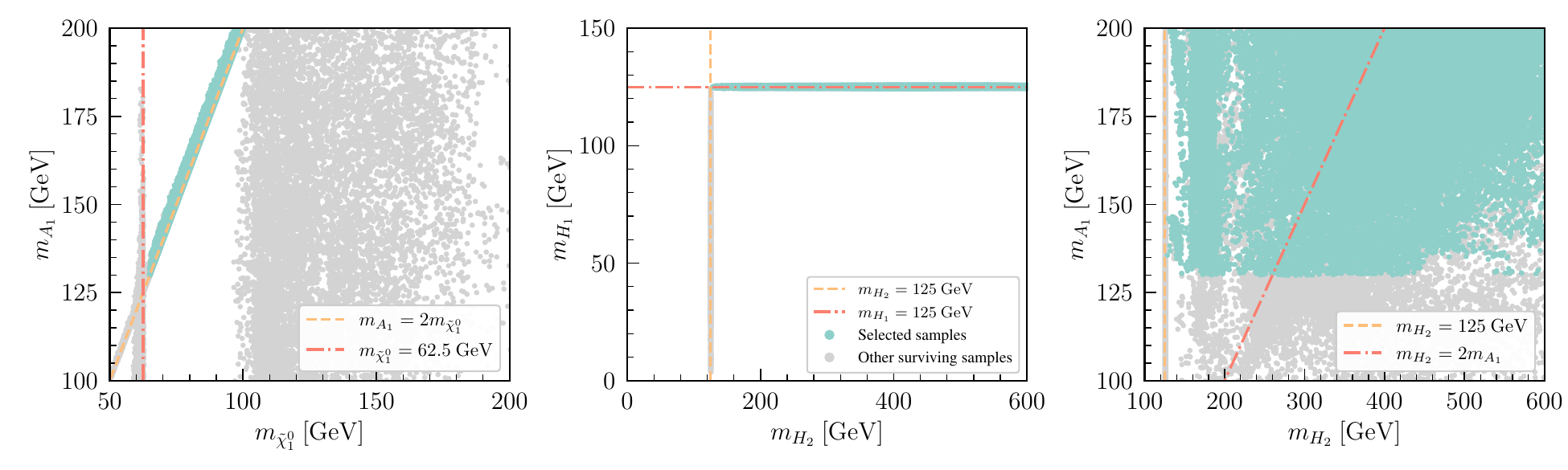}
    \caption{Surviving samples in the $m_{\tilde{\chi}_1^0}$--$m_{A_1}$ plane (left), the $m_{H_2}$--$m_{H_1}$ plane (middle), and the $m_{H_2}$--$m_{A_1}$ plane (right). Cyan samples satisfy $65~\mathrm{GeV}<m_{\tilde{\chi}_1^0}<100~\mathrm{GeV}$ and $m_{A_1}>2m_{\tilde{\chi}_1^0}$, with $H_1$ identified as the observed SM-like Higgs boson; the remaining surviving samples are shown in light gray. In the left panel, the orange dashed and red dash-dotted lines denote $m_{A_1}=2m_{\tilde{\chi}_1^0}$ and $m_{\tilde{\chi}_1^0}=62.5~\mathrm{GeV}$, respectively. In the middle panel, they denote $m_{H_2}=125~\mathrm{GeV}$ and $m_{H_1}=125~\mathrm{GeV}$, respectively. In the right panel, they denote $m_{H_2}=125~\mathrm{GeV}$ and $m_{H_2}=2m_{A_1}$, respectively. The on-shell decay $H_2\to A_1A_1$ is allowed below the red dash-dotted line in the right panel.}
    \label{f1}
\end{figure*}

We explore the parameter space using NMSSMTools~6.2.0~\cite{Ellwanger:2004xm,Ellwanger:2005dv,Ellwanger:2006rn}. An initial random scan is performed with the 11 input parameters sampled over the following ranges:
\begin{equation}
\begin{aligned}
M_0 &\in [1,5]~\mathrm{TeV},\\
M_1,\ M_2 &\in [-1,1]~\mathrm{TeV},\\
M_3 &\in [2,5]~\mathrm{TeV},\\
A_0 &\in [-5,5]~\mathrm{TeV},\\
A_\lambda,\ A_\kappa &\in [-1,1]~\mathrm{TeV},\\
\mu_{\mathrm{eff}} &\in [100,300]~\mathrm{GeV},\\
\tan\beta &\in [1,30],\\
\lambda &\in [0.1,0.6],\\
\kappa &\in [-0.6,0.6].
\end{aligned}
\end{equation}
We subsequently use Markov chain Monte Carlo (MCMC) sampling to explore the viable regions more densely. During this stage, the parameters are allowed to extend beyond the initial random-scan ranges. The candidate samples are subjected to the theoretical and experimental constraints and the subsequent checks described in Sec.~\ref{sec:model}. The surviving samples from this two-stage procedure are used in the following analysis.

Figure~\ref{f1} shows the surviving samples in the $m_{\tilde{\chi}_1^0}$--$m_{A_1}$ plane (left), the $m_{H_2}$--$m_{H_1}$ plane (middle), and the $m_{H_2}$--$m_{A_1}$ plane (right). In the left panel, surviving samples with $m_{\tilde{\chi}_1^0}<100~\mathrm{GeV}$ are predominantly concentrated near $m_{\tilde{\chi}_1^0}\approx m_{A_1}/2$ or $m_{\tilde{\chi}_1^0}\approx62.5~\mathrm{GeV}$, corresponding to the $A_1$ and SM-like Higgs funnel regions, respectively. These two regions lie near the orange dashed and red dash-dotted lines, which indicate the corresponding resonance conditions.

The middle panel displays two branches corresponding to the identification of either $H_1$ or $H_2$ with the observed SM-like Higgs boson. When $H_2$ is SM-like, the lighter state $H_1$ can span a broad mass range, from very low masses up to nearly $125~\mathrm{GeV}$. In the right panel, the red dash-dotted line marks the threshold $m_{H_2}=2m_{A_1}$. A substantial subset of the surviving samples lies below this line, where $m_{H_2}>2m_{A_1}$, demonstrating that the on-shell decay $H_2\to A_1A_1$ is kinematically allowed over an appreciable region of the surviving parameter space.

We highlight in cyan the samples satisfying $65~\mathrm{GeV}<m_{\tilde{\chi}_1^0}<100~\mathrm{GeV}$ and $m_{A_1}>2m_{\tilde{\chi}_1^0}$, with $H_1$ identified as the observed SM-like Higgs boson; the remaining surviving samples are shown in light gray. The subsequent phenomenological analysis focuses on these cyan samples. We study $H_2$ as the parent state of the cascade $H_2\to A_1A_1$ when $m_{H_2}>2m_{A_1}$.

\begin{figure*}[!ht]
    \centering
    \includegraphics[width=\linewidth]{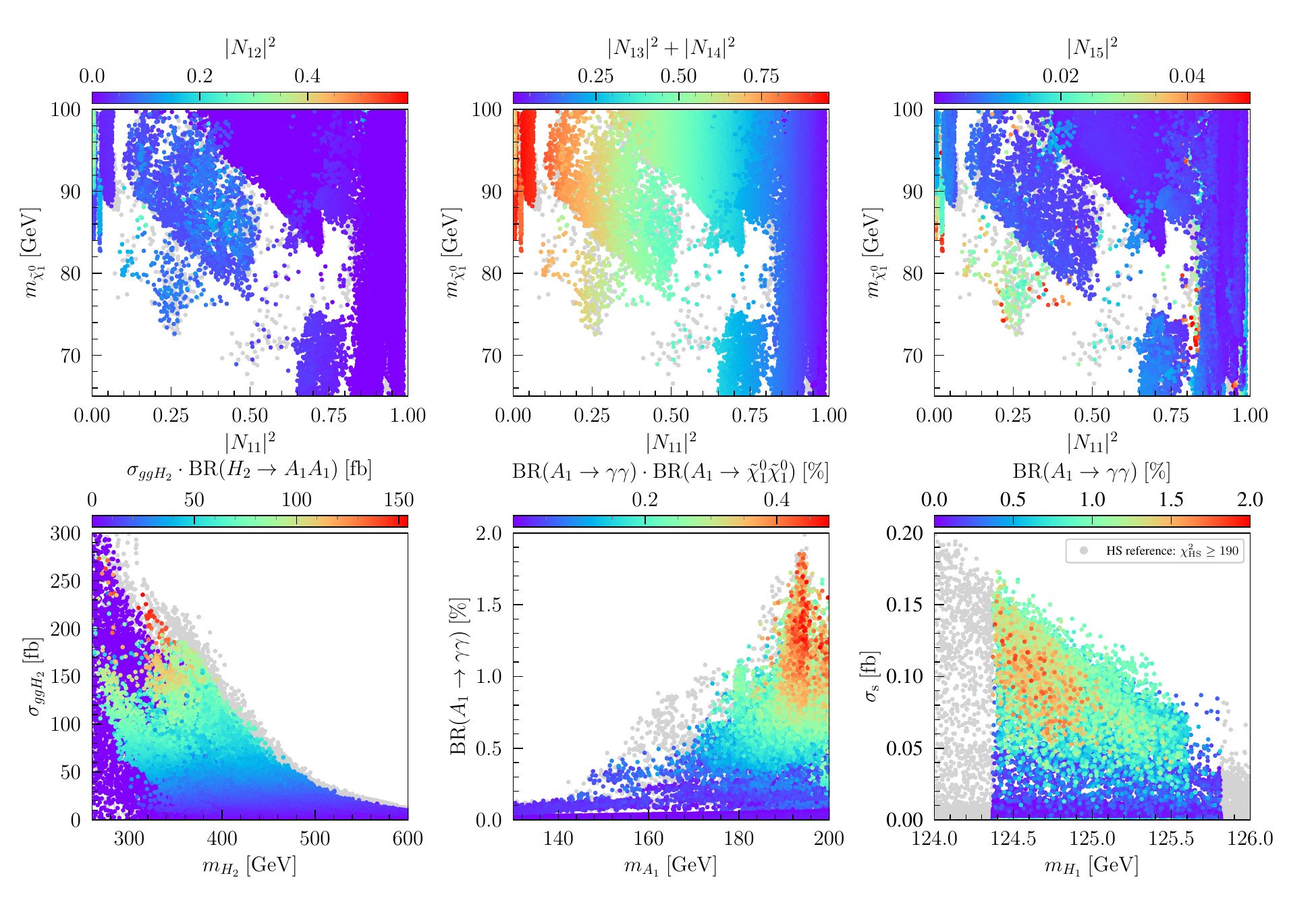}
    \vspace{-20pt}
    \caption{Colored points denote the selected surviving samples highlighted in cyan in Fig.~\ref{f1}. Light-gray points represent reference samples with $\chi^2_{\mathrm{HS}}\geq190$, shown for comparison. Upper panels: the lightest-neutralino mass $m_{\tilde{\chi}_1^0}$ versus the bino fraction $|N_{11}|^2$, with colors indicating the wino fraction $|N_{12}|^2$ (left), the higgsino fraction $|N_{13}|^2+|N_{14}|^2$ (middle), and the singlino fraction $|N_{15}|^2$ (right). Lower left: the gluon-fusion production cross section $\sigma_{ggH_2}$ versus $m_{H_2}$, with colors indicating $\sigma_{ggH_2}\mathrm{BR}(H_2\to A_1A_1)$. Lower middle: $\mathrm{BR}(A_1\to\gamma\gamma)$ versus $m_{A_1}$, with colors indicating $\mathrm{BR}(A_1\to\gamma\gamma)\mathrm{BR}(A_1\to\tilde{\chi}_1^0\tilde{\chi}_1^0)$. Lower right: the cascade signal cross section $\sigma_{\mathrm{s}}$ versus the SM-like Higgs mass $m_{H_1}$, with colors indicating $\mathrm{BR}(A_1\to\gamma\gamma)$. All collider cross sections are evaluated at $\sqrt{s}=14~\mathrm{TeV}$.}
    \label{f2}
\end{figure*}

\begin{figure*}[!htb]
    \centering
    \includegraphics[width=\linewidth]{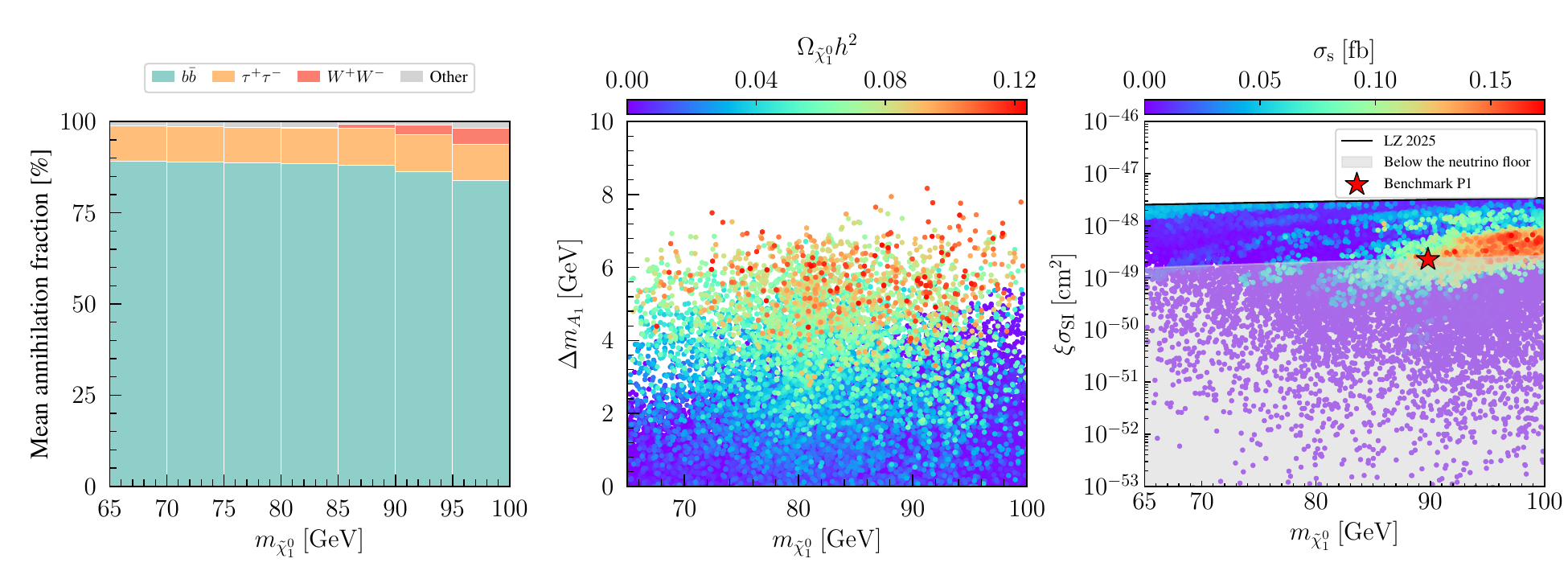}
    \vspace{-20pt}
    \caption{Left: mean fractional contributions of individual channels to low-velocity neutralino annihilation in 5~GeV bins of $m_{\tilde{\chi}_1^0}$, averaged over the selected surviving samples in each bin. The category ``Other'' includes the remaining channels, such as $ZZ$, $\gamma\gamma$, $Z\gamma$, $c\bar{c}$, and $gg$. Middle: the samples in the $m_{\tilde{\chi}_1^0}$–$\Delta m_{A_1}$ plane, where $\Delta m_{A_1}\equiv m_{A_1}-2m_{\tilde{\chi}_1^0}$ is expressed in GeV. The colors indicate the neutralino relic density $\Omega h^2$. Right: the relic-density-rescaled spin-independent neutralino–nucleon scattering cross section $\xi\sigma_{\mathrm{SI}}$ versus $m_{\tilde{\chi}_1^0}$, where $\xi=\Omega h^2/0.1188$. The colors indicate the signal cross section $\sigma_{\mathrm{s}}$ at $\sqrt{s}=14~\mathrm{TeV}$ for $gg\to H_2\to A_1A_1\to\gamma\gamma\tilde{\chi}_1^0\tilde{\chi}_1^0$. The black curve denotes the LZ 2025 upper limit~\cite{LZ:2024zvo}, while the gray shaded region indicates the region below the adopted neutrino-floor curve~\cite{Billard:2013qya}.
}
    \label{f3}
\end{figure*}

The upper panels of Fig.~\ref{f2} show the selected samples in the plane of the lightest-neutralino mass $m_{\tilde{\chi}_1^0}$ and the bino fraction $|N_{11}|^2$, with colors indicating the wino fraction $|N_{12}|^2$ (left), the higgsino fraction $|N_{13}|^2+|N_{14}|^2$ (middle), and the singlino fraction $|N_{15}|^2$ (right). 
Among the selected samples, the lower end of the lightest-neutralino mass range generally shifts to higher values as the bino fraction decreases.
Bino-dominated neutralinos with $|N_{11}|^2>0.8$ are found throughout the mass range of 65--100~GeV. 
For states composed almost entirely of bino and higgsino, with negligible wino and singlino admixtures, a bino fraction below 0.5 is found only for $m_{\tilde{\chi}_1^0}\gtrsim90~\mathrm{GeV}$. 
Lower neutralino masses can occur for a given bino fraction when wino and singlino admixtures are present.
For samples with a wino fraction of approximately 0.1 and a negligible singlino component, neutralinos with $|N_{11}|^2\approx0.5$ can be as light as approximately 78~GeV. 
With an additional singlino fraction of approximately 0.02--0.05, viable samples extend down to $m_{\tilde{\chi}_1^0}\approx 73~\mathrm{GeV}$, even with a bino fraction of about 0.25 and a higgsino fraction of about 0.6.

The lower panels of Fig.~\ref{f2} illustrate the production and decay properties relevant to the cascade signal for the selected surviving samples. 
The lower-left panel shows the gluon-fusion production cross section $\sigma_{ggH_2}$ as a function of $m_{H_2}$, with the colors indicating $\sigma_{ggH_2}\cdot\mathrm{BR}(H_2\to A_1A_1)$. The lower-middle panel displays $\mathrm{BR}(A_1\to\gamma\gamma)$ as a function of $m_{A_1}$, with the colors representing the product $\mathrm{BR}(A_1\to\gamma\gamma)\cdot\mathrm{BR}(A_1\to\tilde{\chi}_1^0\tilde{\chi}_1^0)$. 
The lower-right panel shows the signal cross section $\sigma_s$ as a function of the SM-like Higgs mass $m_{H_1}$, with the colors indicating $\mathrm{BR}(A_1\to\gamma\gamma)$. 
At $\sqrt{s}=14~\mathrm{TeV}$, we estimate the gluon-fusion production cross section by rescaling the SM-coupling reference cross section for a Higgs boson of the same mass, taken from the CERN YR4 tables for BSM applications~\cite{deFlorian:2016spz}. 
The reference cross sections are evaluated at NNLO+NNLL QCD accuracy in the narrow-width approximation. 
We use linear interpolation between the tabulated masses and do not apply the separately provided electroweak correction factors:
\begin{equation}
\sigma_{ggH_2}\approx \left.\sigma_{\mathrm{SM}}^{\mathrm{NNLO+NNLL}}(gg\to h)\right|_{m_h=m_{H_2}}\,|C_{ggH_2}|^2,
\end{equation}
where $C_{ggH_2}$ is the effective coupling of $H_2$ to gluons normalized to that of an SM Higgs boson with the same mass. In the narrow-width approximation, the cross section for $gg\to H_2\to A_1A_1\to\gamma\gamma+\tilde{\chi}_1^0\tilde{\chi}_1^0$ is
\begin{equation}
\begin{aligned}
\sigma_s ={}& 2\,\sigma_{ggH_2}\,\cdot\mathrm{BR}(H_2\to A_1A_1)\\
&\times\mathrm{BR}(A_1\to\gamma\gamma)\,\cdot\mathrm{BR}(A_1\to\tilde{\chi}_1^0\tilde{\chi}_1^0).
\end{aligned}
\end{equation}
The factor of two accounts for the two possible assignments of the diphoton and invisible decays to the two identical $A_1$ bosons.

As shown in the lower-left panel of Fig.~\ref{f2}, lighter $H_2$ states tend to have larger gluon-fusion production cross sections $\sigma_{ggH_2}$ within our scan.
However, the selected samples satisfy $m_{A_1}>2m_{\tilde{\chi}_1^0}>130~\mathrm{GeV}$, and the on-shell decay $H_2\to A_1A_1$ is kinematically forbidden for $m_{H_2}<2m_{A_1}$. 
Samples below this threshold, therefore, do not contribute to the on-shell cascade signal, despite potentially large $H_2$ production cross sections.
After including the branching fraction for $H_2\to A_1A_1$, the cross section $\sigma_{ggH_2}\cdot\mathrm{BR}(H_2\to A_1A_1)$ reaches approximately 150~fb near $m_{H_2}=350~\mathrm{GeV}$, where the corresponding gluon-fusion production cross section is approximately 200~fb. 
For $m_{H_2}>400~\mathrm{GeV}$, the largest gluon-fusion production cross sections among the surviving samples decrease rapidly, falling to a few tens of femtobarns as $m_{H_2}$ approaches 500~GeV.

The lower-middle panel of Fig.~\ref{f2} shows that larger diphoton branching fractions are preferentially obtained for samples with $m_{A_1}\approx195~\mathrm{GeV}$, where $\mathrm{BR}(A_1\to\gamma\gamma)$ approaches 2\%. 
Although the diphoton and invisible decay modes compete with each other, both branching fractions can be enhanced when the partial widths of the dominant fermionic decay modes, particularly $A_1\to b\bar{b}$, are suppressed.
Their product, $\mathrm{BR}(A_1\to\gamma\gamma)\cdot\mathrm{BR}(A_1\to\tilde{\chi}_1^0\tilde{\chi}_1^0)$, approaches 0.5\% in the same pseudoscalar mass region.
As shown in the lower-right panel, the signal cross section can reach approximately 0.17~fb for $m_{H_1}\approx124.4~\mathrm{GeV}$. 
Within the selected surviving samples, larger signal cross sections tend to occur at lower values of $m_{H_1}$ and are associated with larger diphoton branching fractions of $A_1$.

Figure~\ref{f3} summarizes the neutralino annihilation channels, relic density, and direct detection predictions for the selected surviving samples. 
The left panel shows the mean fractional contributions of individual channels to low-velocity neutralino annihilation in 5~GeV bins of $m_{\tilde{\chi}_1^0}$. 
For each channel, its fractional contribution is averaged over the selected surviving samples in the corresponding mass bin. The $b\bar{b}$, $\tau^+\tau^-$, and $W^+W^-$ channels are shown individually, while the remaining channels, including $ZZ$, $\gamma\gamma$, $Z\gamma$, $c\bar{c}$, and $gg$, are grouped into ``Other''.
The middle panel shows the samples in the $m_{\tilde{\chi}_1^0}$--$\Delta m_{A_1}$ plane, where $\Delta m_{A_1}\equiv m_{A_1}-2m_{\tilde{\chi}_1^0}$, with the colors indicating the neutralino relic density $\Omega_{\tilde{\chi}_1^0}h^2$. 
The right panel displays the relic-density-rescaled spin-independent neutralino--nucleon scattering cross section $\xi\sigma_{\mathrm{SI}}$ as a function of $m_{\tilde{\chi}_1^0}$, where $\xi=\Omega_{\tilde{\chi}_1^0}h^2/0.1188$. 
The colors indicate the cascade signal cross section $\sigma_s$ at $\sqrt{s}=14~\mathrm{TeV}$. 
The black curve denotes the LZ 2025 upper limit~\cite{LZ:2024zvo}, while the gray shaded region indicates the region below the adopted neutrino-floor curve~\cite{Billard:2013qya}.

The left panel of Fig.~\ref{f3} shows that $b\bar{b}$ provides the largest mean contribution throughout the neutralino mass range considered, decreasing from approximately 89\% in the 65--70~GeV bin to approximately 84\% in the 95--100~GeV bin. 
The mean $\tau^+\tau^-$ contribution remains close to 10\%. 
The $W^+W^-$ contribution increases in the higher-mass bins, reaching approximately 4.3\% in the 95--100~GeV bin.
The middle panel illustrates the dependence of the neutralino relic density on the mass difference $\Delta m_{A_1}$. 
Within our scan, samples in which the lightest neutralino accounts for all of the dark matter and reproduces the observed relic density tend to have $\Delta m_{A_1}\gtrsim3~\mathrm{GeV}$.
Smaller positive mass splittings $\Delta m_{A_1}$ can enhance neutralino annihilation through the $A_1$ resonance, thereby reducing the thermal relic density. 
Assuming that the local neutralino fraction follows its cosmological abundance, the direct-detection event rates are correspondingly suppressed by $\xi=\Omega_{\tilde{\chi}_1^0}h^2/0.1188$.
This abundance suppression can help these samples satisfy direct-detection constraints, with the neutralino constituting only part of the dark matter.
The right panel demonstrates the complementarity between the cascade search and dark matter direct detection. 
Among the samples below the LZ exclusion curve, the signal cross section can reach approximately 0.17~fb, and values of about 0.16~fb are obtained even below the adopted neutrino floor. 
The sample marked by the red star is chosen as the benchmark point for the Monte Carlo analysis in the following section.

Table~\ref{tab:benchmarks} lists the input parameters and relevant observables for four representative benchmark points, including their neutralino compositions, relic densities, direct detection cross sections, and cascade signal rates at $\sqrt{s}=14~\mathrm{TeV}$. 
P1 and P2 illustrate scenarios with a subdominant neutralino dark matter component and a kinematically allowed $H_2\to A_1A_1$ decay. 
P3 and P4 reproduce the observed dark matter relic density, but satisfy $m_{H_2}<2m_{A_1}$, so the on-shell cascade channel is closed and the corresponding signal cross section vanishes.

\begin{table}[!htb]
\centering
\caption{Input parameters and relevant observables for four benchmark points.
Units and numerical scale factors are specified in the first column;
each tabulated value is to be multiplied by the indicated scale factor.
Nonzero values are rounded to four significant figures.
Collider production cross sections are evaluated at
$\sqrt{s}=14\,\mathrm{TeV}$.
The input parameters follow the scale conventions specified in
Sec.~\ref{sec:model}.
Here $\xi=\Omega_{\tilde{\chi}_1^0}h^2/0.1188$ and
$\Delta m_{A_1}=m_{A_1}-2m_{\tilde{\chi}_1^0}$,
where $m_{\tilde{\chi}_1^0}$ denotes the positive physical mass.
$|\mathcal P_{12}|^2$ denotes the singlet fraction of $A_1$
in the physical doublet--singlet basis.
$R_{\max}$ denotes the largest ratio of the predicted signal rate
to the corresponding experimental upper limit reported by SModelS.}
\label{tab:benchmarks}
\begingroup
\small
\setlength{\tabcolsep}{3pt}
\renewcommand{\arraystretch}{1.05}
\begin{tabular}{@{}lcccc@{}}
\toprule
\toprule
Parameter & P1 & P2 & P3 & P4 \\
\midrule
$M_0\ [\mathrm{GeV}]$ & $3105$ & $2442$ & $3671$ & $4067$ \\
$M_1\ [\mathrm{GeV}]$ & $-141.3$ & $-172.2$ & $-111.6$ & $-140.0$ \\
$M_2\ [\mathrm{GeV}]$ & $1805$ & $1365$ & $1103$ & $1243$ \\
$M_3\ [\mathrm{GeV}]$ & $3586$ & $2894$ & $3326$ & $4267$ \\
$A_0\ [\mathrm{GeV}]$ & $-142.5$ & $-173.6$ & $-4744$ & $-4398$ \\
$A_\lambda\ [\mathrm{GeV}]$ & $157.1$ & $182.1$ & $-89.71$ & $-60.47$ \\
$A_\kappa\ [\mathrm{GeV}]$ & $274.4$ & $203.4$ & $918.1$ & $914.5$ \\
$\mu_{\mathrm{eff}}\ [\mathrm{GeV}]$ & $175.4$ & $176.5$ & $747.7$ & $754.7$ \\
$\tan\beta$ & $9.659$ & $10.16$ & $2.380$ & $2.502$ \\
$\lambda$ & $0.4602$ & $0.4317$ & $0.4998$ & $0.4875$ \\
$\kappa$ & $0.5738$ & $0.5602$ & $-0.03290$ & $-0.03815$ \\
\addlinespace
$m_{H_1}\ [\mathrm{GeV}]$ & $124.8$ & $124.4$ & $125.2$ & $125.2$ \\
$m_{H_2}\ [\mathrm{GeV}]$ & $415.7$ & $436.0$ & $166.3$ & $163.7$ \\
$m_{A_1}\ [\mathrm{GeV}]$ & $179.6$ & $194.1$ & $151.0$ & $191.4$ \\
$m_{\tilde{\chi}_1^0}\ [\mathrm{GeV}]$ & $89.80$ & $97.02$ & $72.44$ & $92.48$ \\
$m_{\tilde{\chi}_1^\pm}\ [\mathrm{GeV}]$ & $180.8$ & $180.8$ & $731.5$ & $751.6$ \\
$\Delta m_{A_1}\ [\mathrm{GeV}]$ & $0.02741$ & $0.02401$ & $6.176$ & $6.411$ \\
\addlinespace
$|N_{11}|^2$ & $0.9188$ & $0.9090$ & $0.9662$ & $0.9686$ \\
$|N_{12}|^2\ [10^{-5}]$ & $1.029$ & $2.998$ & $0.2898$ & $0.1756$ \\
$|N_{13}|^2\ [10^{-2}]$ & $7.131$ & $7.739$ & $0.3525$ & $0.3418$ \\
$|N_{14}|^2\ [10^{-3}]$ & $8.225$ & $12.17$ & $0.005504$ & $0.01317$ \\
$|N_{15}|^2\ [10^{-2}]$ & $0.1640$ & $0.1431$ & $3.024$ & $2.796$ \\
$|\mathcal{P}_{12}|^2$ & $0.9999$ & $0.9999$ & $0.9980$ & $0.9983$ \\
\addlinespace
$\Omega_{\tilde{\chi}_1^0}h^2\ [10^{-2}]$ & $0.04610$ & $0.04598$ & $11.88$ & $12.03$ \\
$\sigma_{\mathrm{SI}}\ [10^{-48}\,\mathrm{cm}^{2}]$ & $58.20$ & $170.7$ & $0.04608$ & $0.01524$ \\
$\xi\sigma_{\mathrm{SI}}\ [10^{-49}\,\mathrm{cm}^{2}]$ & $2.258$ & $6.606$ & $0.4609$ & $0.1543$ \\
\addlinespace
$\sigma_{ggH_2}\ [\mathrm{fb}]$ & $37.29$ & $26.74$ & $1064$ & $1224$ \\
$\sigma_{\mathrm{s}}\ [\mathrm{fb}]$ & $0.1400$ & $0.1733$ & $0$ & $0$ \\
$\chi^2_{\mathrm{HS}}$ & $169.4$ & $185.5$ & $161.8$ & $162.7$ \\
$R_{\max}$ & $0.4482$ & $0.03050$ & $0.06311$ & $0.03180$ \\
\bottomrule
\bottomrule
\end{tabular}
\endgroup
\end{table}

\section{Collider analysis at the HL-LHC}
\label{sec:MC}

\begin{figure}[!htb]
\centering
\includegraphics[width=0.8\linewidth]{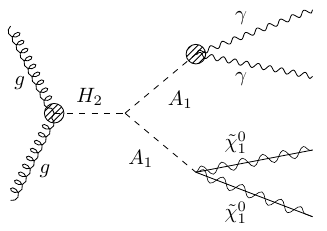}
\caption{\label{feyn} A representative diagram for the signal process $gg\to H_2\to A_1A_1\to\gamma\gamma+\tilde{\chi}_1^0\tilde{\chi}_1^0$. The hatched circles represent the loop-induced effective $ggH_2$ and $A_1\gamma\gamma$ vertices.}
\end{figure}

We investigate the diphoton plus missing transverse momentum signature at the HL-LHC with a center-of-mass energy of $\sqrt{s}=14~\mathrm{TeV}$ and an integrated luminosity of $\mathcal{L}=3000~\mathrm{fb}^{-1}$. 
The signal arises from gluon-fusion production of the next-to-lightest CP-even Higgs boson $H_2$, followed by the cascade decay $H_2\to A_1A_1$, with one $A_1$ decaying into two photons and the other into a pair of lightest neutralinos. 
The neutralinos escape the detector and give rise to missing transverse momentum. 
A representative diagram for the signal process is shown in Fig.~\ref{feyn}, where the hatched circles denote the loop-induced effective $ggH_2$ and $A_1\gamma\gamma$ vertices.

The SM backgrounds considered in this analysis include resonant Higgs production and nonresonant photon production. The resonant contribution arises from the decay $h\to\gamma\gamma$ of the SM Higgs boson, produced through gluon fusion, association with a top-quark pair, or vector-boson fusion (VBF). 
These contributions are collectively denoted by $h_{\gamma\gamma}$. 
The nonresonant backgrounds include diphoton production in association with a vector boson or a jet, denoted by $V\gamma\gamma$ and $\gamma\gamma j$, respectively, where $V=W^\pm,Z$. 
We also consider the single-photon processes $V\gamma$ and $\gamma j$, retaining events that satisfy the two-photon reconstruction and selection requirements after showering and detector simulation.
These backgrounds can acquire missing transverse momentum from neutrinos in the final state or from detector effects.

Monte Carlo simulations of the signal and background processes are performed using MadGraph5\_aMC@NLO~3.6.2~\cite{Alwall:2011uj,Alwall:2014hca}, with the NNPDF23LO1 parton distribution functions~\cite{Ball:2013hta}. 
For the nonresonant backgrounds, matrix-element samples with an additional jet are included and matched to the parton shower using the MLM prescription~\cite{Alwall:2007fs,Alwall:2008qv}, with the matrix-element separation parameter set to $\texttt{xqcut}=25~\mathrm{GeV}$. 
Particle decays, parton showering, and hadronization are handled by PYTHIA~8.2~\cite{Sjostrand:2014zea}, interfaced with MadGraph5\_aMC@NLO through MG5aMC\_PY8\_interface. Detector effects are simulated using Delphes~3.5.0 with the default CMS detector card~\cite{deFavereau:2013fsa,Selvaggi:2014mya}, and jets are reconstructed using the anti-$k_T$ algorithm~\cite{Cacciari:2008gp} implemented in FastJet~\cite{Cacciari:2011ma}. 
The simulated events are analyzed using MadAnalysis~5 v1.9.60~\cite{Conte:2012fm}.

We generate $10^5$ signal events and $10^6$ events for each background process at $\sqrt{s}=14~\mathrm{TeV}$.
As a basic selection, we require exactly two reconstructed photons and veto events containing reconstructed electrons or muons, denoted by $\ell=e,\mu$, to suppress backgrounds with leptonic decays of vector bosons. 
Both photons are required to satisfy $|\eta_\gamma|<2.37$. These requirements define the basic selection:
\begin{equation}
N(\ell)=0,\qquad N(\gamma)=2,\qquad |\eta_\gamma|<2.37.
\end{equation}

\begin{figure*}[!htb]
\centering
\includegraphics[width=\linewidth]{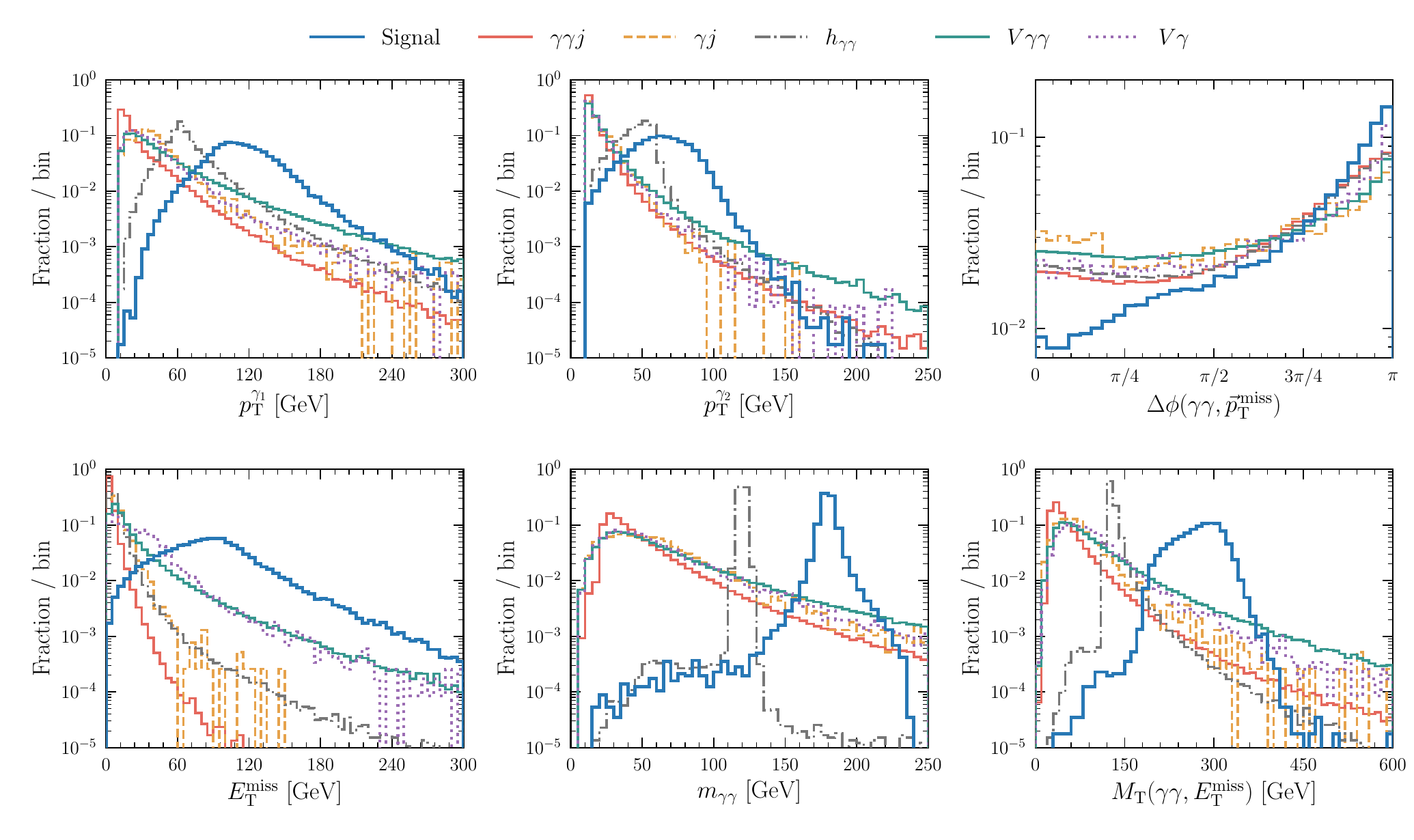}
\caption{Normalized kinematic distributions for the signal and SM backgrounds after the basic selection.
Upper panels: the leading-photon transverse momentum $p_{\mathrm{T}}^{\gamma_1}$ (left), the subleading-photon transverse momentum $p_{\mathrm{T}}^{\gamma_2}$ (middle), and the azimuthal separation $\Delta\phi(\gamma\gamma,\vec{p}_{\mathrm{T}}^{\,\mathrm{miss}})$ (right).
Lower panels: the missing transverse momentum $E_{\mathrm{T}}^{\mathrm{miss}}$ (left), the diphoton invariant mass $m_{\gamma\gamma}$ (middle), and the transverse mass $M_{\mathrm{T}}(\gamma\gamma,E_{\mathrm{T}}^{\mathrm{miss}})$ (right).
Each distribution is normalized to the total number of events passing the basic selection in the corresponding sample, including events outside the displayed range.}
\label{f4}
\end{figure*}

Figure~\ref{f4} shows the normalized kinematic distributions of the signal and SM backgrounds after the basic selection, before any additional kinematic requirements are imposed.
The upper panels display the leading-photon transverse momentum $p_{\mathrm{T}}^{\gamma_1}$ (left), the subleading-photon transverse momentum $p_{\mathrm{T}}^{\gamma_2}$ (middle), and the azimuthal separation $\Delta\phi(\gamma\gamma,\vec{p}_{\mathrm{T}}^{\,\mathrm{miss}})$ between the diphoton system and the missing transverse momentum (right).
The lower panels show the missing transverse momentum $E_{\mathrm{T}}^{\mathrm{miss}}$ (left), the diphoton invariant mass $m_{\gamma\gamma}$ (middle), and the transverse mass $M_{\mathrm{T}}(\gamma\gamma,E_{\mathrm{T}}^{\mathrm{miss}})$ (right).
The photons are ordered by transverse momentum, with $p_{\mathrm{T}}^{\gamma_1}\geq p_{\mathrm{T}}^{\gamma_2}$.
For each sample, the bin contents are normalized to the total number of events passing the basic selection, including events outside the displayed ranges.

\begin{table*}[!htb]
\centering
\caption{Cumulative selection efficiencies for benchmark point P1 and the SM backgrounds at the 14~TeV HL-LHC. All efficiencies are normalized to the corresponding initial event samples. The two photon transverse-momentum requirements are reported as a single selection stage. Entries of zero indicate that no simulated events survive the corresponding selection and do not imply a vanishing physical background.}
\label{t01}
\small
\setlength{\tabcolsep}{4pt}
\renewcommand{\arraystretch}{1.15}
\begin{tabular}{@{}lcccccc@{}}
\toprule
\toprule
\multirow{2}{*}{Selection} & \multirow{2}{*}{Signal} & \multicolumn{5}{c}{Backgrounds} \\
\cmidrule(lr){3-7}
& & $\gamma\gamma j$ & $\gamma j$ & $h_{\gamma\gamma}$ & $V\gamma\gamma$ & $V\gamma$ \\
\midrule
Initial sample
& 1.000 & 1.000 & 1.000 & 1.000 & 1.000 & 1.000 \\
Basic selection
& 0.569 & 0.599 & $3.87\times10^{-3}$ & 0.661 & 0.434 & $1.17\times10^{-2}$ \\
$p_{\mathrm{T}}^{\gamma_1}>65~\mathrm{GeV},\quad p_{\mathrm{T}}^{\gamma_2}>50~\mathrm{GeV}$
& 0.417 & $1.34\times10^{-2}$ & $1.00\times10^{-4}$ & 0.118 & $2.89\times10^{-2}$ & $4.21\times10^{-4}$ \\
$\Delta\phi(\gamma\gamma,\vec{p}_{\mathrm{T}}^{\,\mathrm{miss}})>\pi/6$
& 0.397 & $1.15\times10^{-2}$ & $8.20\times10^{-5}$ & 0.104 & $2.48\times10^{-2}$ & $3.71\times10^{-4}$ \\
$E_{\mathrm{T}}^{\mathrm{miss}}>70~\mathrm{GeV}$
& 0.267 & $4.90\times10^{-5}$ & 0 & $1.41\times10^{-3}$ & $3.75\times10^{-3}$ & $6.50\times10^{-5}$ \\
$174.63~\mathrm{GeV}\le m_{\gamma\gamma}\le184.63~\mathrm{GeV}$
& 0.194 & 0 & 0 & 0 & $1.29\times10^{-4}$ & 0 \\
$M_{\mathrm{T}}(\gamma\gamma,E_{\mathrm{T}}^{\mathrm{miss}})<400~\mathrm{GeV}$
& 0.194 & 0 & 0 & 0 & $1.03\times10^{-4}$ & 0 \\
\bottomrule
\bottomrule
\end{tabular}
\end{table*}

The upper-left and upper-middle panels show that the signal photons typically have larger transverse momenta than those in the nonresonant backgrounds.
Requiring both photons to have sufficiently large transverse momenta therefore suppresses these backgrounds while retaining a substantial fraction of the signal.
We impose
\begin{equation}
p_{\mathrm{T}}^{\gamma_1}>65~\mathrm{GeV},\qquad p_{\mathrm{T}}^{\gamma_2}>50~\mathrm{GeV}.
\end{equation}
The two requirements are reported as a single selection stage in Table~\ref{t01}.

The upper-right panel displays $\Delta\phi(\gamma\gamma,\vec{p}_{\mathrm{T}}^{\,\mathrm{miss}})$, which characterizes the transverse recoil between the visible and invisible systems.
For the signal process $H_2\to A_1A_1$, one $A_1$ decays into two photons, while the other decays invisibly into $\tilde{\chi}_1^0\tilde{\chi}_1^0$.
In the limit of negligible transverse momentum of $H_2$ and ideal reconstruction, the visible and invisible systems balance each other in the transverse plane, yielding $\Delta\phi(\gamma\gamma,\vec{p}_{\mathrm{T}}^{\,\mathrm{miss}})\approx\pi$.
Initial-state radiation and detector effects broaden this configuration.
Since several backgrounds also populate the large-$\Delta\phi$ region, this observable alone provides only limited separation.
We impose the loose requirement
\begin{equation}
\Delta\phi(\gamma\gamma,\vec{p}_{\mathrm{T}}^{\,\mathrm{miss}})>\frac{\pi}{6},
\end{equation}
which removes configurations with a small azimuthal separation while retaining most of the signal surviving the photon transverse-momentum requirements.

The lower-left panel of Fig.~\ref{f4} shows that the signal has a broader missing-transverse-momentum distribution, with substantial contributions at larger $E_{\mathrm{T}}^{\mathrm{miss}}$ than the backgrounds dominated by instrumental missing momentum.
The $V\gamma\gamma$ and $V\gamma$ samples also contain contributions with genuine missing transverse momentum from neutrinos in vector-boson decays.
Following the photon transverse-momentum and azimuthal-separation requirements, we impose
\begin{equation}
E_{\mathrm{T}}^{\mathrm{miss}}>70~\mathrm{GeV}.
\end{equation}
This requirement strongly suppresses backgrounds with small missing transverse momentum.

The lower-middle panel shows a pronounced signal peak near $m_{\gamma\gamma}=179.63~\mathrm{GeV}$, corresponding to $m_{A_1}$ for benchmark point P1.
The $h_{\gamma\gamma}$ background instead peaks near $125~\mathrm{GeV}$, whereas the nonresonant backgrounds exhibit continuous diphoton invariant-mass distributions and tend to populate the lower-mass region.
We therefore require
\begin{equation}
\left|m_{\gamma\gamma}-179.63~\mathrm{GeV}\right|\leq5~\mathrm{GeV},
\end{equation}
corresponding to $174.63~\mathrm{GeV}\leq m_{\gamma\gamma}\leq184.63~\mathrm{GeV}$.
This mass window selects the $A_1$ resonance while strongly suppressing the SM Higgs contribution and the continuum backgrounds.

The lower-right panel displays the transverse mass of the diphoton system and the missing transverse momentum.
Following the \texttt{MT\_MET} definition implemented in MadAnalysis~5, we use
\begin{equation}
\begin{aligned}
M_{\mathrm{T}}^2(\gamma\gamma,E_{\mathrm{T}}^{\mathrm{miss}})
={}&\left(E_{\mathrm{T}}^{\gamma\gamma}+E_{\mathrm{T}}^{\mathrm{miss}}\right)^2\\
&-\left|\vec{p}_{\mathrm{T}}^{\,\gamma\gamma}+\vec{p}_{\mathrm{T}}^{\,\mathrm{miss}}\right|^2,
\end{aligned}
\end{equation}
where
\begin{equation}
E_{\mathrm{T}}^{\gamma\gamma}=\sqrt{m_{\gamma\gamma}^2+\left|\vec{p}_{\mathrm{T}}^{\,\gamma\gamma}\right|^2},
\qquad
\vec{p}_{\mathrm{T}}^{\,\gamma\gamma}=\vec{p}_{\mathrm{T}}^{\,\gamma_1}+\vec{p}_{\mathrm{T}}^{\,\gamma_2},
\end{equation}
and $E_{\mathrm{T}}^{\mathrm{miss}}=|\vec{p}_{\mathrm{T}}^{\,\mathrm{miss}}|$.
This definition assigns zero trial mass to the invisible system; it does not assume that the neutralinos are massless.
As the final selection, we require
\begin{equation}
M_{\mathrm{T}}(\gamma\gamma,E_{\mathrm{T}}^{\mathrm{miss}})<400~\mathrm{GeV}.
\end{equation}
This requirement can further suppress the $V\gamma$ and $V\gamma\gamma$ backgrounds, whose distributions exhibit long tails extending into the region around $400~\mathrm{GeV}$.
Table~\ref{t01} summarizes the cumulative selection efficiencies at each stage, normalized to the corresponding initial event samples.

The basic selection efficiently suppresses the single-photon backgrounds, rejecting approximately $99.6\%$ and $98.8\%$ of the initial $\gamma j$ and $V\gamma$ samples, respectively. 
The photon transverse-momentum requirements further suppress the nonresonant backgrounds, whose photons generally have softer transverse-momentum distributions than those of the signal, while retaining $41.7\%$ of the initial signal sample. 
The subsequent azimuthal-separation requirement retains approximately $95.2\%$ of the signal surviving the photon transverse-momentum requirements.
The missing-transverse-momentum requirement strongly suppresses the $\gamma\gamma j$ and $h_{\gamma\gamma}$ backgrounds, yielding a cumulative signal efficiency of $26.7\%$.
The diphoton mass window further reduces the backgrounds while retaining $19.4\%$ of the initial signal sample, leaving $V\gamma\gamma$ as the only background with a nonzero efficiency in the simulated samples.
The final transverse-mass requirement reduces the remaining $V\gamma\gamma$ background while retaining nearly all of the surviving signal.
After all selections, the cumulative efficiencies are $19.4\%$ for the signal and $1.03\times10^{-4}$ for the $V\gamma\gamma$ background.

We estimate the median expected discovery significance using the Asimov formula~\cite{Cowan:2010js},
\begin{equation}
Z=\sqrt{2\left[(S+B)\ln\left(1+\frac{S}{B}\right)-S\right]},
\end{equation}
where $S$ and $B$ are the expected signal and total background event yields after all selections, respectively. 
The $V\gamma\gamma$ background is normalized to a cross section of 0.4925~pb before the analysis selections, as calculated using MadGraph5\_aMC@NLO~\cite{Alwall:2011uj,Alwall:2014hca}. 
With a cumulative selection efficiency of $1.03\times10^{-4}$, it contributes approximately 152.2 background events at an integrated luminosity of $3000~\mathrm{fb}^{-1}$. 
For benchmark point P1, the signal cross section of 0.14~fb and the cumulative efficiency of 0.194 correspond to approximately 81.6 signal events. 
Neglecting systematic uncertainties and taking $V\gamma\gamma$ as the remaining background, we obtain an expected significance of $6.13\sigma$ at the 14~TeV HL-LHC.

Beyond benchmark point P1, the selected surviving samples include scenarios near the neutrino floor with signal cross sections reaching approximately 0.17~fb, exceeding the P1 value of 0.14~fb. 
These larger production rates motivate extending the diphoton plus missing transverse momentum analysis to additional benchmark points, with the event selection adapted to their Higgs masses and kinematic distributions. 
The result for P1 illustrates the potential of the cascade channel $pp\to H_2\to A_1A_1\to\gamma\gamma+\tilde{\chi}_1^0\tilde{\chi}_1^0$ to probe dark matter scenarios below the neutrino floor at the HL-LHC.

\section{Conclusions}
\label{sec:conclusion}
In this work, we have investigated the potential of Higgs cascade decays to probe neutralino dark matter below the neutrino floor in the semi-constrained NMSSM with nonuniversal gaugino masses at the GUT scale. 
Focusing on lightest-neutralino masses between 65 and 100~GeV, we have explored the parameter space subject to theoretical and experimental constraints and examined the neutralino composition, relic density, and direct detection predictions. 
We have studied the cascade $pp\to H_2\to A_1A_1\to\gamma\gamma+\tilde{\chi}_1^0\tilde{\chi}_1^0$, where $H_1$ is the observed SM-like Higgs boson and $H_2$ is the parent state. 
Monte Carlo simulations of a representative benchmark point and the relevant SM backgrounds have been performed to estimate the sensitivity of this channel at the 14~TeV HL-LHC.

Within the scanned parameter space, surviving samples with neutralino masses below 100~GeV are found near the $A_1$ or SM-like Higgs resonance. 
Among the selected surviving samples, bino-dominated LSPs with $|N_{11}|^2>0.8$ occur throughout the neutralino mass range of 65--100~GeV. The minimum neutralino mass generally increases as the bino fraction decreases. 
For states composed almost entirely of bino and higgsino, a bino fraction below 0.5 is found only for $m_{\tilde{\chi}_1^0}\gtrsim90~\mathrm{GeV}$. 
With a wino admixture, samples with $|N_{11}|^2\approx0.5$ extend down to approximately $78~\mathrm{GeV}$, while an additional small singlino component allows viable samples near 73~GeV with a bino fraction of approximately 0.25.
The sample-averaged contributions to low-velocity neutralino annihilation are dominated by $b\bar{b}$, with $\tau^+\tau^-$ contributing approximately 10\% and $W^+W^-$ remaining subdominant.
Within our scan, samples in which the lightest neutralino accounts for all of the dark matter and reproduces the observed relic density tend to have $\Delta m_{A_1}\gtrsim3~\mathrm{GeV}$.
Samples with smaller positive mass differences can have a reduced relic density because of resonantly enhanced annihilation and remain viable when neutralinos constitute only part of the dark matter.

The cross section $\sigma_{ggH_2}\cdot\mathrm{BR}(H_2\to A_1A_1)$ reaches approximately 150~fb near $m_{H_2}=350~\mathrm{GeV}$, while $\mathrm{BR}(A_1\to\gamma\gamma)$ approaches 2\% for $m_{A_1}\approx 195~\mathrm{GeV}$. 
The resulting diphoton plus missing transverse momentum signal cross section can reach approximately 0.17~fb near the neutrino floor, with values of about 0.16~fb obtained even for samples whose relic-density-rescaled spin-independent scattering cross sections lie below it.
For the representative benchmark point P1, the Monte Carlo analysis yields an expected statistical significance of $6.13\sigma$ at the 14~TeV HL-LHC with an integrated luminosity of $3000~\mathrm{fb}^{-1}$, neglecting systematic uncertainties.
These results illustrate the potential of Higgs cascade decays to probe neutralino dark matter scenarios below the neutrino floor and provide a complementary search channel to direct detection.

\acknowledgments
We thank Yifan Zhu for helpful discussions. 
This work was supported by the National Natural Science Foundation of China under Grant No. 12275066 and by the startup research funds of Henan University. 
The work of H. Yang was also supported by the National Natural Science Foundation of China under Grant number: W2441004.
And the work of K. Wang was also supported by the Open Project of the Shanghai Key Laboratory for Particle Physics and Cosmology under Grant No. 22DZ2229013-3.

\appendix


\bibliographystyle{apsrev4-1}
\bibliography{apssamp}

@article{LZ:2024zvo,
    author = "Aalbers, J. and others",
    collaboration = "LZ",
    title = "{Dark Matter Search Results from 4.2{\,}{\,}Tonne-Years of Exposure of the LUX-ZEPLIN (LZ) Experiment}",
    eprint = "2410.17036",
    archivePrefix = "arXiv",
    primaryClass = "hep-ex",
    reportNumber = "FERMILAB-PUB-24-0796-V",
    doi = "10.1103/4dyc-z8zf",
    journal = "Phys. Rev. Lett.",
    volume = "135",
    number = "1",
    pages = "011802",
    year = "2025"
}

@article{Billard:2013qya,
    author = "Billard, J. and Strigari, L. and Figueroa-Feliciano, E.",
    title = "{Implication of neutrino backgrounds on the reach of next generation dark matter direct detection experiments}",
    eprint = "1307.5458",
    archivePrefix = "arXiv",
    primaryClass = "hep-ph",
    doi = "10.1103/PhysRevD.89.023524",
    journal = "Phys. Rev. D",
    volume = "89",
    number = "2",
    pages = "023524",
    year = "2014"
}

@article{Arcadi:2017kky,
    author = "Arcadi, Giorgio and Dutra, Ma{\'\i}ra and Ghosh, Pradipta and Lindner, Manfred and Mambrini, Yann and Pierre, Mathias and Profumo, Stefano and Queiroz, Farinaldo S.",
    title = "{The waning of the WIMP? A review of models, searches, and constraints}",
    eprint = "1703.07364",
    archivePrefix = "arXiv",
    primaryClass = "hep-ph",
    doi = "10.1140/epjc/s10052-018-5662-y",
    journal = "Eur. Phys. J. C",
    volume = "78",
    number = "3",
    pages = "203",
    year = "2018"
}

@article{Boveia:2018yeb,
    author = "Boveia, Antonio and Doglioni, Caterina",
    title = "{Dark Matter Searches at Colliders}",
    eprint = "1810.12238",
    archivePrefix = "arXiv",
    primaryClass = "hep-ex",
    doi = "10.1146/annurev-nucl-101917-021008",
    journal = "Ann. Rev. Nucl. Part. Sci.",
    volume = "68",
    pages = "429--459",
    year = "2018"
}

@article{ATLAS:2021kxv,
    author = "Aad, Georges and others",
    collaboration = "ATLAS",
    title = "{Search for new phenomena in events with an energetic jet and missing transverse momentum in $pp$ collisions at $\sqrt {s}$ =13  TeV with the ATLAS detector}",
    eprint = "2102.10874",
    archivePrefix = "arXiv",
    primaryClass = "hep-ex",
    reportNumber = "CERN-EP-2020-238",
    doi = "10.1103/PhysRevD.103.112006",
    journal = "Phys. Rev. D",
    volume = "103",
    number = "11",
    pages = "112006",
    year = "2021"
}

@article{ATLAS:2020uiq,
    author = "Aad, Georges and others",
    collaboration = "ATLAS",
    title = "{Search for dark matter in association with an energetic photon in $pp$ collisions at $\sqrt{s}$ = 13 TeV with the ATLAS detector}",
    eprint = "2011.05259",
    archivePrefix = "arXiv",
    primaryClass = "hep-ex",
    reportNumber = "CERN-EP-2020-178",
    doi = "10.1007/JHEP02(2021)226",
    journal = "JHEP",
    volume = "02",
    pages = "226",
    year = "2021"
}

@article{Abdallah:2015ter,
    author = "Abdallah, Jalal and others",
    title = "{Simplified Models for Dark Matter Searches at the LHC}",
    eprint = "1506.03116",
    archivePrefix = "arXiv",
    primaryClass = "hep-ph",
    reportNumber = "FERMILAB-PUB-15-283-CD, CERN-PH-TH-2015-139",
    doi = "10.1016/j.dark.2015.08.001",
    journal = "Phys. Dark Univ.",
    volume = "9-10",
    pages = "8--23",
    year = "2015"
}

@article{Baum:2017enm,
    author = "Baum, Sebastian and Carena, Marcela and Shah, Nausheen R. and Wagner, Carlos E. M.",
    title = "{Higgs portals for thermal Dark Matter. EFT perspectives and the NMSSM}",
    eprint = "1712.09873",
    archivePrefix = "arXiv",
    primaryClass = "hep-ph",
    reportNumber = "NORDITA-2017-130, FERMILAB-PUB-17-611-T, EFI-17-25, WSU-HEP-1715",
    doi = "10.1007/JHEP04(2018)069",
    journal = "JHEP",
    volume = "04",
    pages = "069",
    year = "2018"
}

@article{ATLAS:2021jbf,
    author = "Aad, Georges and others",
    collaboration = "ATLAS",
    title = "{Search for dark matter in events with missing transverse momentum and a Higgs boson decaying into two photons in pp collisions at $ \sqrt{s} $ = 13 TeV with the ATLAS detector}",
    eprint = "2104.13240",
    archivePrefix = "arXiv",
    primaryClass = "hep-ex",
    reportNumber = "CERN-EP-2021-012",
    doi = "10.1007/JHEP10(2021)013",
    journal = "JHEP",
    volume = "10",
    pages = "013",
    year = "2021"
}

@article{Jungman:1995df,
    author = "Jungman, Gerard and Kamionkowski, Marc and Griest, Kim",
    title = "{Supersymmetric dark matter}",
    eprint = "hep-ph/9506380",
    archivePrefix = "arXiv",
    reportNumber = "SU-4240-605, UCSD-PTH-95-02, IASSNS-HEP-95-14, CU-TP-677",
    doi = "10.1016/0370-1573(95)00058-5",
    journal = "Phys. Rept.",
    volume = "267",
    pages = "195--373",
    year = "1996"
}

@article{Djouadi:2005gj,
    author = "Djouadi, Abdelhak",
    title = "{The Anatomy of electro-weak symmetry breaking. II. The Higgs bosons in the minimal supersymmetric model}",
    eprint = "hep-ph/0503173",
    archivePrefix = "arXiv",
    reportNumber = "LPT-ORSAY-05-18",
    doi = "10.1016/j.physrep.2007.10.005",
    journal = "Phys. Rept.",
    volume = "459",
    pages = "1--241",
    year = "2008"
}

@article{Cao:2012im,
    author = "Cao, Jun-Jie and Heng, Zhaoxia and Yang, Jin Min and Zhu, Jingya",
    title = "{Higgs decay to dark matter in low energy SUSY: is it detectable at the LHC ?}",
    eprint = "1203.0694",
    archivePrefix = "arXiv",
    primaryClass = "hep-ph",
    doi = "10.1007/JHEP06(2012)145",
    journal = "JHEP",
    volume = "06",
    pages = "145",
    year = "2012"
}

@article{ATLAS:2023tkt,
    author = "Aad, Georges and others",
    collaboration = "ATLAS",
    title = "{Combination of searches for invisible decays of the Higgs boson using 139 fb{\ensuremath{-}}1 of proton-proton collision data at s=13 TeV collected with the ATLAS experiment}",
    eprint = "2301.10731",
    archivePrefix = "arXiv",
    primaryClass = "hep-ex",
    reportNumber = "CERN-EP-2022-289",
    doi = "10.1016/j.physletb.2023.137963",
    journal = "Phys. Lett. B",
    volume = "842",
    pages = "137963",
    year = "2023"
}

@article{Baer:2012uy,
    author = "Baer, Howard and Barger, Vernon and Huang, Peisi and Tata, Xerxes",
    title = "{Natural Supersymmetry: LHC, dark matter and ILC searches}",
    eprint = "1203.5539",
    archivePrefix = "arXiv",
    primaryClass = "hep-ph",
    reportNumber = "UH-511-1190-12",
    doi = "10.1007/JHEP05(2012)109",
    journal = "JHEP",
    volume = "05",
    pages = "109",
    year = "2012"
}

@article{Cao:2018rix,
    author = "Cao, Junjie and He, Yangle and Shang, Liangliang and Zhang, Yang and Zhu, Pengxuan",
    title = "{Current status of a natural NMSSM in light of LHC 13 TeV data and XENON-1T results}",
    eprint = "1810.09143",
    archivePrefix = "arXiv",
    primaryClass = "hep-ph",
    reportNumber = "CoEPP-MN-18-8",
    doi = "10.1103/PhysRevD.99.075020",
    journal = "Phys. Rev. D",
    volume = "99",
    number = "7",
    pages = "075020",
    year = "2019"
}

@article{ATLAS:2020syg,
    author = "Aad, Georges and others",
    collaboration = "ATLAS",
    title = "{Search for squarks and gluinos in final states with jets and missing transverse momentum using 139 fb$^{-1}$ of $\sqrt{s}$ =13 TeV $pp$ collision data with the ATLAS detector}",
    eprint = "2010.14293",
    archivePrefix = "arXiv",
    primaryClass = "hep-ex",
    reportNumber = "CERN-EP-2020-166",
    doi = "10.1007/JHEP02(2021)143",
    journal = "JHEP",
    volume = "02",
    pages = "143",
    year = "2021"
}

@article{CMS:2019zmd,
    author = "A. M. Sirunyan and others",
    collaboration = "CMS",
    title = "{Search for supersymmetry in proton-proton collisions at 13 TeV in final states with jets and missing transverse momentum}",
    eprint = "1908.04722",
    archivePrefix = "arXiv",
    primaryClass = "hep-ex",
    reportNumber = "CMS-SUS-19-006, CERN-EP-2019-152",
    doi = "10.1007/JHEP10(2019)244",
    journal = "JHEP",
    volume = "10",
    pages = "244",
    year = "2019"
}

@article{King:2014xwa,
    author = {King, S. F. and M{\"u}hlleitner, M. and Nevzorov, R. and Walz, K.},
    title = "{Discovery Prospects for NMSSM Higgs Bosons at the High-Energy Large Hadron Collider}",
    eprint = "1408.1120",
    archivePrefix = "arXiv",
    primaryClass = "hep-ph",
    doi = "10.1103/PhysRevD.90.095014",
    journal = "Phys. Rev. D",
    volume = "90",
    number = "9",
    pages = "095014",
    year = "2014"
}

@article{Ellwanger:2017skc,
    author = "Ellwanger, Ulrich and Rodriguez-Vazquez, Matias",
    title = "{Simultaneous search for extra light and heavy Higgs bosons via cascade decays}",
    eprint = "1707.08522",
    archivePrefix = "arXiv",
    primaryClass = "hep-ph",
    reportNumber = "LPT-ORSAY-17-33",
    doi = "10.1007/JHEP11(2017)008",
    journal = "JHEP",
    volume = "11",
    pages = "008",
    year = "2017"
}

@article{Baum:2019uzg,
    author = "Baum, Sebastian and Shah, Nausheen R. and Freese, Katherine",
    title = "{The NMSSM is within Reach of the LHC: Mass Correlations {\textbackslash}{\&} Decay Signatures}",
    eprint = "1901.02332",
    archivePrefix = "arXiv",
    primaryClass = "hep-ph",
    reportNumber = "NORDITA-2018-128, LCTP-18-32, WSU-HEP-1901",
    doi = "10.1007/JHEP04(2019)011",
    journal = "JHEP",
    volume = "04",
    pages = "011",
    year = "2019"
}

@article{Ellwanger:2022jtd,
    author = "Ellwanger, Ulrich and Hugonie, Cyril",
    title = "{Benchmark planes for Higgs-to-Higgs decays in the NMSSM}",
    eprint = "2203.05049",
    archivePrefix = "arXiv",
    primaryClass = "hep-ph",
    reportNumber = "LUPM 22-004",
    doi = "10.1140/epjc/s10052-022-10364-3",
    journal = "Eur. Phys. J. C",
    volume = "82",
    number = "5",
    pages = "406",
    year = "2022"
}

@article{Baum:2017gbj,
    author = "Baum, Sebastian and Freese, Katherine and Shah, Nausheen R. and Shakya, Bibhushan",
    title = "{NMSSM Higgs boson search strategies at the LHC and the mono-Higgs signature in particular}",
    eprint = "1703.07800",
    archivePrefix = "arXiv",
    primaryClass = "hep-ph",
    reportNumber = "NORDITA-2017-24, MCTP-17-03, WSU-HEP-1702",
    doi = "10.1103/PhysRevD.95.115036",
    journal = "Phys. Rev. D",
    volume = "95",
    number = "11",
    pages = "115036",
    year = "2017"
}

@article{Ellwanger:2009dp,
    author = "Ellwanger, Ulrich and Hugonie, Cyril and Teixeira, Ana M.",
    title = "{The Next-to-Minimal Supersymmetric Standard Model}",
    eprint = "0910.1785",
    archivePrefix = "arXiv",
    primaryClass = "hep-ph",
    reportNumber = "LPT-ORSAY-09-76, CFTP-09-032, LPTA-09-066",
    doi = "10.1016/j.physrep.2010.07.001",
    journal = "Phys. Rept.",
    volume = "496",
    pages = "1--77",
    year = "2010"
}

@article{Miller:2003ay,
    author = "Miller, D. J. and Nevzorov, R. and Zerwas, P. M.",
    title = "{The Higgs sector of the next-to-minimal supersymmetric standard model}",
    eprint = "hep-ph/0304049",
    archivePrefix = "arXiv",
    reportNumber = "CERN-TH-2003-077, DESY-03-066, ITEP-5-03",
    doi = "10.1016/j.nuclphysb.2003.12.021",
    journal = "Nucl. Phys. B",
    volume = "681",
    pages = "3--30",
    year = "2004"
}

@article{Cao:2011re,
    author = "Cao, Jun-Jie and Hikasa, Ken-ichi and Wang, Wenyu and Yang, Jin Min ",
    title = "{Light dark matter in NMSSM and implication on Higgs phenomenology}",
    eprint = "1104.1754",
    archivePrefix = "arXiv",
    primaryClass = "hep-ph",
    doi = "10.1016/j.physletb.2011.07.086",
    journal = "Phys. Lett. B",
    volume = "703",
    pages = "292--297",
    year = "2011"
}

@article{Li:2023kbf,
    author = "Li, Weichao and Qiao, Haoxue and Wang, Kun and Zhu, Jingya",
    title = "{Light dark matter confronted with the 95 GeV diphoton excess}",
    eprint = "2312.17599",
    archivePrefix = "arXiv",
    primaryClass = "hep-ph",
    month = "12",
    year = "2023"
}

@article{Li:2025qkg,
    author = "Li, Fei and Cao, Junjie",
    title = "{Natural realization of tens-of-GeV dark matter in the general NMSSM}",
    eprint = "2511.06329",
    archivePrefix = "arXiv",
    primaryClass = "hep-ph",
    doi = "10.1103/x7j7-5mtq",
    journal = "Phys. Rev. D",
    volume = "113",
    number = "11",
    pages = "115002",
    year = "2026"
}

@article{Cao:2013gba,
    author = "Cao, Junjie and Ding, Fangfang and Han, Chengcheng and Yang, Jin Min and Zhu, Jingya",
    title = "{A light Higgs scalar in the NMSSM confronted with the latest LHC Higgs data}",
    eprint = "1309.4939",
    archivePrefix = "arXiv",
    primaryClass = "hep-ph",
    doi = "10.1007/JHEP11(2013)018",
    journal = "JHEP",
    volume = "11",
    pages = "018",
    year = "2013"
}

@article{Ma:2020mjz,
    author = "Ma, Shiquan and Wang, Kun and Zhu, Jingya",
    title = "{Higgs decay to light (pseudo)scalars in the semi-constrained NMSSM}",
    eprint = "2006.03527",
    archivePrefix = "arXiv",
    primaryClass = "hep-ph",
    reportNumber = "WHU-HEP-PH-TEV008",
    doi = "10.1088/1674-1137/abce4f",
    journal = "Chin. Phys. C",
    volume = "45",
    number = "2",
    pages = "023113",
    year = "2021"
}

@article{Guchait:2016pes,
    author = "Guchait, Monoranjan and Kumar, Jacky",
    title = "{Diphoton Signal of light pseudoscalar in NMSSM at the LHC}",
    eprint = "1608.05693",
    archivePrefix = "arXiv",
    primaryClass = "hep-ph",
    doi = "10.1103/PhysRevD.95.035036",
    journal = "Phys. Rev. D",
    volume = "95",
    number = "3",
    pages = "035036",
    year = "2017"
}

@article{Gunion:2005rw,
    author = "Gunion, John F. and Hooper, Dan and McElrath, Bob",
    title = "{Light neutralino dark matter in the NMSSM}",
    eprint = "hep-ph/0509024",
    archivePrefix = "arXiv",
    reportNumber = "UCD-2005-06, FERMILAB-PUB-05-350-A",
    doi = "10.1103/PhysRevD.73.015011",
    journal = "Phys. Rev. D",
    volume = "73",
    pages = "015011",
    year = "2006"
}

@article{Wang:2018vrr,
    author = "Wang, Fei and Wang, Kun and Yang, Jin Min and Zhu, Jingya",
    title = "{Solving the muon g-2 anomaly in CMSSM extension with non-universal gaugino masses}",
    eprint = "1808.10851",
    archivePrefix = "arXiv",
    primaryClass = "hep-ph",
    doi = "10.1007/JHEP12(2018)041",
    journal = "JHEP",
    volume = "12",
    pages = "041",
    year = "2018"
}

@article{Dong:2024juh,
    author = "Dong, Yabo and Wang, Kun and Yuan, Hailong and Zhu, Jingya and Zhu, Pengxuan",
    title = "{Revisiting CMSSM with non-universal gaugino masses under current constraints}",
    eprint = "2412.20003",
    archivePrefix = "arXiv",
    primaryClass = "hep-ph",
    doi = "10.1007/JHEP03(2025)207",
    journal = "JHEP",
    volume = "03",
    pages = "207",
    year = "2025"
}

@article{Du:2017str,
    author = "Du, Xiaokang and Wang, Fei",
    title = "{NMSSM From Alternative Deflection in Generalized Deflected Anomaly Mediated SUSY Breaking}",
    eprint = "1710.06105",
    archivePrefix = "arXiv",
    primaryClass = "hep-ph",
    doi = "10.1140/epjc/s10052-018-5921-y",
    journal = "Eur. Phys. J. C",
    volume = "78",
    number = "5",
    pages = "431",
    year = "2018"
}

@article{Du:2018pko,
    author = "Du, Xiao Kang and Liu, Guo-Li and Wang, Fei and Wang, Wenyu and Yang, Jin Min and Zhang, Yang",
    title = "{NMSSM with generalized deflected mirage mediation}",
    eprint = "1804.07335",
    archivePrefix = "arXiv",
    primaryClass = "hep-ph",
    doi = "10.1140/epjc/s10052-019-6903-4",
    journal = "Eur. Phys. J. C",
    volume = "79",
    number = "5",
    pages = "397",
    year = "2019"
}

@article{Lian:2024smg,
    author = "Lian, Jingwei",
    title = "{95~GeV excesses in the $Z_3$-symmetric next-to-minimal supersymmetric standard model}",
    eprint = "2406.10969",
    archivePrefix = "arXiv",
    primaryClass = "hep-ph",
    doi = "10.1103/PhysRevD.110.115018",
    journal = "Phys. Rev. D",
    volume = "110",
    number = "11",
    pages = "115018",
    year = "2024"
}

@article{Ellwanger:2024txc,
    author = "Ellwanger, Ulrich and Hugonie, Cyril",
    title = "{NMSSM with correct relic density and an additional 95~GeV Higgs boson}",
    eprint = "2403.16884",
    archivePrefix = "arXiv",
    primaryClass = "hep-ph",
    doi = "10.1140/epjc/s10052-024-12886-4",
    journal = "Eur. Phys. J. C",
    volume = "84",
    number = "5",
    pages = "526",
    year = "2024"
}

@article{Cao:2023gkc,
    author = "Cao, Junjie and Jia, Xinglong and Lian, Jingwei and Meng, Lei",
    title = "{95~GeV diphoton and $b\bar{b}$ excesses in the general next-to-minimal supersymmetric standard model}",
    eprint = "2310.08436",
    archivePrefix = "arXiv",
    primaryClass = "hep-ph",
    doi = "10.1103/PhysRevD.109.075001",
    journal = "Phys. Rev. D",
    volume = "109",
    number = "7",
    pages = "075001",
    year = "2024"
}

@article{Zhou:2021pit,
    author = "Zhou, Haijing and Cao, Junjie and Lian, Jingwei and Zhang, Di",
    title = "{Singlino-dominated dark matter in $Z_3$-symmetric NMSSM}",
    eprint = "2102.05309",
    archivePrefix = "arXiv",
    primaryClass = "hep-ph",
    doi = "10.1103/PhysRevD.104.015017",
    journal = "Phys. Rev. D",
    volume = "104",
    number = "1",
    pages = "015017",
    year = "2021"
}

@article{Carena:2015moc,
    author = "Carena, Marcela and Haber, Howard E. and Low, Ian and Shah, Nausheen R. and Wagner, Carlos E. M.",
    title = "{Alignment limit of the NMSSM Higgs sector}",
    eprint = "1510.09137",
    archivePrefix = "arXiv",
    primaryClass = "hep-ph",
    reportNumber = "FERMILAB-PUB-15-407-T, EFI-15-32, MCTP-15-15, SCIPP-15-12, WSU-HEP-1505",
    doi = "10.1103/PhysRevD.93.035013",
    journal = "Phys. Rev. D",
    volume = "93",
    number = "3",
    pages = "035013",
    year = "2016"
}

@article{Li:2022etb,
    author = "Li, Weichao and Qiao, Haoxue and Zhu, Jingya",
    title = "{Light Higgs boson in the NMSSM confronted with the CMS di-photon and di-tau excesses*}",
    eprint = "2212.11739",
    archivePrefix = "arXiv",
    primaryClass = "hep-ph",
    doi = "10.1088/1674-1137/acfaf1",
    journal = "Chin. Phys. C",
    volume = "47",
    number = "12",
    pages = "123102",
    year = "2023"
}

@article{Maniatis:2009re,
    author = "Maniatis, M.",
    title = "{The Next-to-Minimal Supersymmetric extension of the Standard Model reviewed}",
    eprint = "0906.0777",
    archivePrefix = "arXiv",
    primaryClass = "hep-ph",
    reportNumber = "HD-THEP-09-9",
    doi = "10.1142/S0217751X10049827",
    journal = "Int. J. Mod. Phys. A",
    volume = "25",
    pages = "3505--3602",
    year = "2010"
}

@article{Cao:2016uwt,
    author = "Cao, Junjie and Guo, Xiaofei and He, Yangle and Wu, Peiwen and Zhang, Yang",
    title = "{Diphoton signal of the light Higgs boson in natural NMSSM}",
    eprint = "1612.08522",
    archivePrefix = "arXiv",
    primaryClass = "hep-ph",
    doi = "10.1103/PhysRevD.95.116001",
    journal = "Phys. Rev. D",
    volume = "95",
    number = "11",
    pages = "116001",
    year = "2017"
}

@article{Ellwanger:2004xm,
    author = "Ellwanger, Ulrich and Gunion, John F. and Hugonie, Cyril",
    title = "{NMHDECAY: A Fortran code for the Higgs masses, couplings and decay widths in the NMSSM}",
    eprint = "hep-ph/0406215",
    archivePrefix = "arXiv",
    reportNumber = "LPT-ORSAY-04-32, UCD-04-22, IFIC-04-33",
    doi = "10.1088/1126-6708/2005/02/066",
    journal = "JHEP",
    volume = "02",
    pages = "066",
    year = "2005"
}

@article{Ellwanger:2005dv,
    author = "Ellwanger, Ulrich and Hugonie, Cyril",
    title = "{NMHDECAY 2.0: An Updated program for sparticle masses, Higgs masses, couplings and decay widths in the NMSSM}",
    eprint = "hep-ph/0508022",
    archivePrefix = "arXiv",
    reportNumber = "LPT-ORSAY-05-52",
    doi = "10.1016/j.cpc.2006.04.004",
    journal = "Comput. Phys. Commun.",
    volume = "175",
    pages = "290--303",
    year = "2006"
}

@article{Ellwanger:2006rn,
    author = "Ellwanger, Ulrich and Hugonie, Cyril",
    title = "{NMSPEC: A Fortran code for the sparticle and Higgs masses in the NMSSM with GUT scale boundary conditions}",
    eprint = "hep-ph/0612134",
    archivePrefix = "arXiv",
    reportNumber = "LPT-ORSAY-06-79, LPTA-MONTPELLIER-06-6",
    doi = "10.1016/j.cpc.2007.05.001",
    journal = "Comput. Phys. Commun.",
    volume = "177",
    pages = "399--407",
    year = "2007"
}

@article{Das:2011dg,
    author = "Das, Debottam and Ellwanger, Ulrich and Teixeira, Ana M.",
    title = "{NMSDECAY: A Fortran Code for Supersymmetric Particle Decays in the Next-to-Minimal Supersymmetric Standard Model}",
    eprint = "1106.5633",
    archivePrefix = "arXiv",
    primaryClass = "hep-ph",
    reportNumber = "LPT-ORSAY-11-56, PCCF-RI-1104",
    doi = "10.1016/j.cpc.2011.11.021",
    journal = "Comput. Phys. Commun.",
    volume = "183",
    pages = "774--779",
    year = "2012"
}

@article{Muhlleitner:2003vg,
    author = "Muhlleitner, M. and Djouadi, A. and Mambrini, Y.",
    title = "{SDECAY: A Fortran code for the decays of the supersymmetric particles in the MSSM}",
    eprint = "hep-ph/0311167",
    archivePrefix = "arXiv",
    reportNumber = "CERN-TH-2003-252, PM-03-22, PSI-PR-03-17",
    doi = "10.1016/j.cpc.2005.01.012",
    journal = "Comput. Phys. Commun.",
    volume = "168",
    pages = "46--70",
    year = "2005"
}

@article{Bechtle:2008jh,
    author = "Bechtle, Philip and Brein, Oliver and Heinemeyer, Sven and Weiglein, Georg and Williams, Karina E.",
    title = "{HiggsBounds: Confronting Arbitrary Higgs Sectors with Exclusion Bounds from LEP and the Tevatron}",
    eprint = "0811.4169",
    archivePrefix = "arXiv",
    primaryClass = "hep-ph",
    reportNumber = "DCPT-08-172, IPPP-08-86, BONN-TH-2008-17",
    doi = "10.1016/j.cpc.2009.09.003",
    journal = "Comput. Phys. Commun.",
    volume = "181",
    pages = "138--167",
    year = "2010"
}

@article{Bechtle:2013xfa,
    author = "Bechtle, Philip and Heinemeyer, Sven and St{\r{a}}l, Oscar and Stefaniak, Tim and Weiglein, Georg",
    title = "{$HiggsSignals$: Confronting arbitrary Higgs sectors with measurements at the Tevatron and the LHC}",
    eprint = "1305.1933",
    archivePrefix = "arXiv",
    primaryClass = "hep-ph",
    reportNumber = "BONN-TH-2013-07, DESY-13-078",
    doi = "10.1140/epjc/s10052-013-2711-4",
    journal = "Eur. Phys. J. C",
    volume = "74",
    number = "2",
    pages = "2711",
    year = "2014"
}

@article{Bahl:2022igd,
    author = {Bahl, Henning and Biek{\"o}tter, Thomas and Heinemeyer, Sven and Li, Cheng and Paasch, Steven and Weiglein, Georg and Wittbrodt, Jonas},
    title = "{HiggsTools: BSM scalar phenomenology with new versions of HiggsBounds and HiggsSignals}",
    eprint = "2210.09332",
    archivePrefix = "arXiv",
    primaryClass = "hep-ph",
    doi = "10.1016/j.cpc.2023.108803",
    journal = "Comput. Phys. Commun.",
    volume = "291",
    pages = "108803",
    year = "2023"
}

@article{CMS:2018szt,
    author = "Sirunyan, A. M. and others",
    collaboration = "CMS",
    title = "{Combined search for electroweak production of charginos and neutralinos in proton-proton collisions at $\sqrt{s} =$ 13 TeV}",
    eprint = "1801.03957",
    archivePrefix = "arXiv",
    primaryClass = "hep-ex",
    reportNumber = "CMS-SUS-17-004, CERN-EP-2017-283",
    doi = "10.1007/JHEP03(2018)160",
    journal = "JHEP",
    volume = "03",
    pages = "160",
    year = "2018"
}

@article{Kraml:2013mwa,
    author = "Kraml, Sabine and Kulkarni, Suchita and Laa, Ursula and Lessa, Andre and Magerl, Wolfgang and Proschofsky-Spindler, Doris and Waltenberger, Wolfgang",
    title = "{SModelS: a tool for interpreting simplified-model results from the LHC and its application to supersymmetry}",
    eprint = "1312.4175",
    archivePrefix = "arXiv",
    primaryClass = "hep-ph",
    doi = "10.1140/epjc/s10052-014-2868-5",
    journal = "Eur. Phys. J. C",
    volume = "74",
    pages = "2868",
    year = "2014"
}

@article{Altakach:2024jwk,
    author = "Altakach, Mohammad Mahdi and Kraml, Sabine and Lessa, Andre and Narasimha, Sahana and Pascal, Timoth{\'e}e and Ramos, Camila and Villamizar, Yoxara and Waltenberger, Wolfgang",
    title = "{SModelS v3: going beyond $ \mathcal{Z} _{2}$ topologies}",
    eprint = "2409.12942",
    archivePrefix = "arXiv",
    primaryClass = "hep-ph",
    doi = "10.1007/JHEP11(2024)074",
    journal = "JHEP",
    volume = "11",
    pages = "074",
    year = "2024"
}

@article{ATLAS:2019lng,
    author = "Aad, Georges and others",
    collaboration = "ATLAS",
    title = "{Searches for electroweak production of supersymmetric particles with compressed mass spectra in $\sqrt{s}=$ 13 TeV $pp$ collisions with the ATLAS detector}",
    eprint = "1911.12606",
    archivePrefix = "arXiv",
    primaryClass = "hep-ex",
    reportNumber = "CERN-EP-2019-242",
    doi = "10.1103/PhysRevD.101.052005",
    journal = "Phys. Rev. D",
    volume = "101",
    number = "5",
    pages = "052005",
    year = "2020"
}

@article{Domingo:2007dx,
    author = "Domingo, Florian and Ellwanger, Ulrich",
    title = "{Updated Constraints from $B$ Physics on the MSSM and the NMSSM}",
    eprint = "0710.3714",
    archivePrefix = "arXiv",
    primaryClass = "hep-ph",
    reportNumber = "LPT-ORSAY-07-88",
    doi = "10.1088/1126-6708/2007/12/090",
    journal = "JHEP",
    volume = "12",
    pages = "090",
    year = "2007"
}

@article{Domingo:2015wyn,
    author = "Domingo, Florian",
    title = "{Update of the flavour-physics constraints in the NMSSM}",
    eprint = "1512.02091",
    archivePrefix = "arXiv",
    primaryClass = "hep-ph",
    reportNumber = "IFT-UAM-CSIC-15-130",
    doi = "10.1140/epjc/s10052-016-4298-z",
    journal = "Eur. Phys. J. C",
    volume = "76",
    number = "8",
    pages = "452",
    year = "2016"
}

@article{Domingo:2008rr,
    author = "Domingo, Florian and Ellwanger, Ulrich and Fullana, Esteban and Hugonie, Cyril and Sanchis-Lozano, Miguel-Angel",
    title = "{Radiative Upsilon decays and a light pseudoscalar Higgs in the NMSSM}",
    eprint = "0810.4736",
    archivePrefix = "arXiv",
    primaryClass = "hep-ph",
    reportNumber = "IFIC-08-53, FTUV-08-1024, LPT-08-81, ANL-HEP-PR-08-64, LPTA-08-059",
    doi = "10.1088/1126-6708/2009/01/061",
    journal = "JHEP",
    volume = "01",
    pages = "061",
    year = "2009"
}

@article{Domingo:2010am,
    author = "Domingo, Florian",
    title = "{Updated constraints from radiative $\Upsilon$ decays on a light CP-odd Higgs}",
    eprint = "1010.4701",
    archivePrefix = "arXiv",
    primaryClass = "hep-ph",
    reportNumber = "TTP10-44, SFB-CPP-10-91",
    doi = "10.1007/JHEP04(2011)016",
    journal = "JHEP",
    volume = "04",
    pages = "016",
    year = "2011"
}

@article{Belanger:2005kh,
    author = "Belanger, G. and Boudjema, F. and Hugonie, C. and Pukhov, A. and Semenov, A.",
    title = "{Relic density of dark matter in the NMSSM}",
    eprint = "hep-ph/0505142",
    archivePrefix = "arXiv",
    doi = "10.1088/1475-7516/2005/09/001",
    journal = "JCAP",
    volume = "09",
    pages = "001",
    year = "2005"
}

@article{Belanger:2008sj,
    author = "Belanger, G. and Boudjema, F. and Pukhov, A. and Semenov, A.",
    title = "{Dark matter direct detection rate in a generic model with micrOMEGAs 2.2}",
    eprint = "0803.2360",
    archivePrefix = "arXiv",
    primaryClass = "hep-ph",
    reportNumber = "LAPTH-1237-08",
    doi = "10.1016/j.cpc.2008.11.019",
    journal = "Comput. Phys. Commun.",
    volume = "180",
    pages = "747--767",
    year = "2009"
}

@article{Alguero:2023zol,
    author = "Alguero, G. and Belanger, G. and Boudjema, F. and Chakraborti, S. and Goudelis, A. and Kraml, S. and Mjallal, A. and Pukhov, A.",
    title = "{micrOMEGAs 6.0: N-component dark matter}",
    eprint = "2312.14894",
    archivePrefix = "arXiv",
    primaryClass = "hep-ph",
    doi = "10.1016/j.cpc.2024.109133",
    journal = "Comput. Phys. Commun.",
    volume = "299",
    pages = "109133",
    year = "2024"
}

@article{Planck:2015fie,
    author = "Ade, P. A. R. and others",
    collaboration = "Planck",
    title = "{Planck 2015 results. XIII. Cosmological parameters}",
    eprint = "1502.01589",
    archivePrefix = "arXiv",
    primaryClass = "astro-ph.CO",
    doi = "10.1051/0004-6361/201525830",
    journal = "Astron. Astrophys.",
    volume = "594",
    pages = "A13",
    year = "2016"
}

@article{LZ:2025igz,
    author = "Akerib, D. S. and others",
    collaboration = "LZ",
    title = "{Searches for Light Dark Matter and Evidence of Coherent Elastic Neutrino-Nucleus Scattering of Solar Neutrinos with the LUX-ZEPLIN (LZ) Experiment}",
    eprint = "2512.08065",
    archivePrefix = "arXiv",
    primaryClass = "hep-ex",
    doi = "10.1103/jvqf-njpj",
    journal = "Phys. Rev. Lett.",
    volume = "137",
    number = "9",
    pages = "091806",
    year = "2026"
}

@article{PandaX:2025rrz,
    author = "Zhang, Minzhen and others",
    collaboration = "PandaX",
    title = "{Search for Light Dark Matter with 259 Days of Data in PandaX-4T}",
    eprint = "2507.11930",
    archivePrefix = "arXiv",
    primaryClass = "hep-ex",
    doi = "10.1103/rtnh-jn8s",
    journal = "Phys. Rev. Lett.",
    volume = "135",
    number = "21",
    pages = "211001",
    year = "2025",
    note = "[Erratum: Phys.Rev.Lett. 136, 069901 (2026)]"
}

@article{XENONCollaborationP:2026ioh,
    author = "Aprile, E. and others",
    collaboration = "XENON",
    title = "{Light Dark Matter Search with 7.8~Tonne-Year of Ionization-Only Data in XENONnT}",
    eprint = "2601.11296",
    archivePrefix = "arXiv",
    primaryClass = "hep-ex",
    doi = "10.1103/2lrq-f6bk",
    journal = "Phys. Rev. Lett.",
    volume = "137",
    number = "5",
    pages = "051003",
    year = "2026"
}

@article{DarkSide-50:2025lns,
    author = "Acerbi, F. and others",
    collaboration = "DarkSide-50, DarkSide-20k",
    title = "{Sensitivity to low-mass WIMPs with an improved liquid argon ionization response model within the DarkSide program}",
    eprint = "2511.13629",
    archivePrefix = "arXiv",
    primaryClass = "hep-ex",
    doi = "10.1103/x3vt-q676",
    journal = "Phys. Rev. D",
    volume = "113",
    number = "12",
    pages = "123045",
    year = "2026"
}

@article{deFlorian:2016spz,
    author = "de Florian, D. and others",
    collaboration = "LHC Higgs Cross Section Working Group",
    title = "{Handbook of LHC Higgs Cross Sections: 4. Deciphering the Nature of the Higgs Sector}",
    eprint = "1610.07922",
    archivePrefix = "arXiv",
    primaryClass = "hep-ph",
    reportNumber = "CERN-2017-002-M, CERN-2017-002",
    doi = "10.23731/CYRM-2017-002",
    journal = "CERN Yellow Rep. Monogr.",
    volume = "2",
    pages = "1--869",
    year = "2017"
}

@article{Alwall:2011uj,
    author = "Alwall, Johan and Herquet, Michel and Maltoni, Fabio and Mattelaer, Olivier and Stelzer, Tim",
    title = "{MadGraph 5 : Going Beyond}",
    eprint = "1106.0522",
    archivePrefix = "arXiv",
    primaryClass = "hep-ph",
    reportNumber = "FERMILAB-PUB-11-448-T",
    doi = "10.1007/JHEP06(2011)128",
    journal = "JHEP",
    volume = "06",
    pages = "128",
    year = "2011"
}

@article{Alwall:2014hca,
    author = "Alwall, J. and Frederix, R. and Frixione, S. and Hirschi, V. and Maltoni, F. and Mattelaer, O. and Shao, H. -S. and Stelzer, T. and Torrielli, P. and Zaro, M.",
    title = "{The automated computation of tree-level and next-to-leading order differential cross sections, and their matching to parton shower simulations}",
    eprint = "1405.0301",
    archivePrefix = "arXiv",
    primaryClass = "hep-ph",
    reportNumber = "CERN-PH-TH-2014-064, CP3-14-18, LPN14-066, MCNET-14-09, ZU-TH-14-14",
    doi = "10.1007/JHEP07(2014)079",
    journal = "JHEP",
    volume = "07",
    pages = "079",
    year = "2014"
}

@article{Ball:2013hta,
    author = "Ball, Richard D. and Bertone, Valerio and Carrazza, Stefano and Del Debbio, Luigi and Forte, Stefano and Guffanti, Alberto and Hartland, Nathan P. and Rojo, Juan",
    collaboration = "NNPDF",
    title = "{Parton distributions with QED corrections}",
    eprint = "1308.0598",
    archivePrefix = "arXiv",
    primaryClass = "hep-ph",
    reportNumber = "EDINBURGH-2013-20, FR-PHENO-2013-008, CERN-PH-TH-2013-075, Edinburgh 2013/20, IFUM-1014-FT, FR-PHENO-2013-008,
  CERN-PH-TH/2013-075",
    doi = "10.1016/j.nuclphysb.2013.10.010",
    journal = "Nucl. Phys. B",
    volume = "877",
    pages = "290--320",
    year = "2013"
}

@article{Alwall:2007fs,
    author = "Alwall, Johan and others",
    title = "{Comparative study of various algorithms for the merging of parton showers and matrix elements in hadronic collisions}",
    eprint = "0706.2569",
    archivePrefix = "arXiv",
    primaryClass = "hep-ph",
    reportNumber = "SLAC-PUB-12604, CERN-PH-TH-2007-066, LU-TP-07-13, KA-TP-06-2007, DCPT-07-62, IPPP-07-31",
    doi = "10.1140/epjc/s10052-007-0490-5",
    journal = "Eur. Phys. J. C",
    volume = "53",
    pages = "473--500",
    year = "2008"
}

@article{Alwall:2008qv,
    author = "Alwall, Johan and de Visscher, Simon and Maltoni, Fabio",
    title = "{QCD radiation in the production of heavy colored particles at the LHC}",
    eprint = "0810.5350",
    archivePrefix = "arXiv",
    primaryClass = "hep-ph",
    doi = "10.1088/1126-6708/2009/02/017",
    journal = "JHEP",
    volume = "02",
    pages = "017",
    year = "2009"
}

@article{Sjostrand:2014zea,
    author = {Sj{\"o}strand, Torbj{\"o}rn and Ask, Stefan and Christiansen, Jesper R. and Corke, Richard and Desai, Nishita and Ilten, Philip and Mrenna, Stephen and Prestel, Stefan and Rasmussen, Christine O. and Skands, Peter Z.},
    title = "{An introduction to PYTHIA 8.2}",
    eprint = "1410.3012",
    archivePrefix = "arXiv",
    primaryClass = "hep-ph",
    reportNumber = "LU-TP-14-36, MCNET-14-22, CERN-PH-TH-2014-190, FERMILAB-PUB-14-316-CD, DESY-14-178, SLAC-PUB-16122",
    doi = "10.1016/j.cpc.2015.01.024",
    journal = "Comput. Phys. Commun.",
    volume = "191",
    pages = "159--177",
    year = "2015"
}

@article{deFavereau:2013fsa,
    author = "de Favereau, J. and Delaere, C. and Demin, P. and Giammanco, A. and Lema{\^\i}tre, V. and Mertens, A. and Selvaggi, M.",
    collaboration = "DELPHES 3",
    title = "{DELPHES 3, A modular framework for fast simulation of a generic collider experiment}",
    eprint = "1307.6346",
    archivePrefix = "arXiv",
    primaryClass = "hep-ex",
    doi = "10.1007/JHEP02(2014)057",
    journal = "JHEP",
    volume = "02",
    pages = "057",
    year = "2014"
}

@article{Selvaggi:2014mya,
    author = "Selvaggi, Michele",
    editor = "Wang, Jianxiong",
    title = "{DELPHES 3: A modular framework for fast-simulation of generic collider experiments}",
    doi = "10.1088/1742-6596/523/1/012033",
    journal = "J. Phys. Conf. Ser.",
    volume = "523",
    pages = "012033",
    year = "2014"
}

@article{Cacciari:2008gp,
    author = "Cacciari, Matteo and Salam, Gavin P. and Soyez, Gregory",
    title = "{The anti-$k_t$ jet clustering algorithm}",
    eprint = "0802.1189",
    archivePrefix = "arXiv",
    primaryClass = "hep-ph",
    reportNumber = "LPTHE-07-03",
    doi = "10.1088/1126-6708/2008/04/063",
    journal = "JHEP",
    volume = "04",
    pages = "063",
    year = "2008"
}

@article{Cacciari:2011ma,
    author = "Cacciari, Matteo and Salam, Gavin P. and Soyez, Gregory",
    title = "{FastJet User Manual}",
    eprint = "1111.6097",
    archivePrefix = "arXiv",
    primaryClass = "hep-ph",
    reportNumber = "CERN-PH-TH-2011-297",
    doi = "10.1140/epjc/s10052-012-1896-2",
    journal = "Eur. Phys. J. C",
    volume = "72",
    pages = "1896",
    year = "2012"
}

@article{Conte:2012fm,
    author = "Conte, Eric and Fuks, Benjamin and Serret, Guillaume",
    title = "{MadAnalysis 5, A User-Friendly Framework for Collider Phenomenology}",
    eprint = "1206.1599",
    archivePrefix = "arXiv",
    primaryClass = "hep-ph",
    reportNumber = "IPHC-PHENO-06",
    doi = "10.1016/j.cpc.2012.09.009",
    journal = "Comput. Phys. Commun.",
    volume = "184",
    pages = "222--256",
    year = "2013"
}

@article{Hall:2011aa,
    author = "Hall, Lawrence J. and Pinner, David and Ruderman, Joshua T.",
    title = "{A Natural SUSY Higgs Near 126 GeV}",
    eprint = "1112.2703",
    archivePrefix = "arXiv",
    primaryClass = "hep-ph",
    reportNumber = "UCB-PTH-11-11",
    doi = "10.1007/JHEP04(2012)131",
    journal = "JHEP",
    volume = "04",
    pages = "131",
    year = "2012"
}

@article{Cao:2012fz,
    author = "Cao, Jun-Jie and Heng, Zhao-Xia and Yang, Jin Min and Zhang, Yan-Ming and Zhu, Jing-Ya",
    title = "{A SM-like Higgs near 125 GeV in low energy SUSY: a comparative study for MSSM and NMSSM}",
    eprint = "1202.5821",
    archivePrefix = "arXiv",
    primaryClass = "hep-ph",
    doi = "10.1007/JHEP03(2012)086",
    journal = "JHEP",
    volume = "03",
    pages = "086",
    year = "2012"
}

@article{King:2012is,
    author = "King, S. F. and Muhlleitner, M. and Nevzorov, R.",
    title = "{NMSSM Higgs Benchmarks Near 125 GeV}",
    eprint = "1201.2671",
    archivePrefix = "arXiv",
    primaryClass = "hep-ph",
    reportNumber = "KA-TP-01-2012, SFB-CPP-12-02, UH511-1188-2012",
    doi = "10.1016/j.nuclphysb.2012.02.010",
    journal = "Nucl. Phys. B",
    volume = "860",
    pages = "207--244",
    year = "2012"
}

@article{Dermisek:2005ar,
    author = "Dermisek, Radovan and Gunion, John F.",
    title = "{Escaping the large fine tuning and little hierarchy problems in the next to minimal supersymmetric model and h ---{\ensuremath{>}} aa decays}",
    eprint = "hep-ph/0502105",
    archivePrefix = "arXiv",
    doi = "10.1103/PhysRevLett.95.041801",
    journal = "Phys. Rev. Lett.",
    volume = "95",
    pages = "041801",
    year = "2005"
}

@article{LZ:2024vge,
    author = "Aalbers, J. and others",
    collaboration = "LZ",
    title = "{Constraints on Covariant Dark-Matter{\textendash}Nucleon Effective Field Theory Interactions from the First Science Run of the LUX-ZEPLIN Experiment}",
    eprint = "2404.17666",
    archivePrefix = "arXiv",
    primaryClass = "hep-ex",
    reportNumber = "FERMILAB-PUB-24-0760-V",
    doi = "10.1103/PhysRevLett.133.221801",
    journal = "Phys. Rev. Lett.",
    volume = "133",
    number = "22",
    pages = "221801",
    year = "2024"
}

@article{LZ:2026axp,
    author = "Akerib, D. S. and others",
    collaboration = "LZ",
    title = "{Search for dark matter particle interactions in an extended nuclear recoil energy window with the LUX-ZEPLIN (LZ) experiment}",
    eprint = "2609.02823",
    archivePrefix = "arXiv",
    primaryClass = "hep-ex",
    month = "9",
    year = "2026"
}

@article{ATLAS:2022hbt,
    author = "Aad, Georges and others",
    collaboration = "ATLAS",
    title = "{Search for direct pair production of sleptons and charginos decaying to two leptons and neutralinos with mass splittings near the W-boson mass in $ \sqrt{s} $ = 13 TeV pp collisions with the ATLAS detector}",
    eprint = "2209.13935",
    archivePrefix = "arXiv",
    primaryClass = "hep-ex",
    reportNumber = "CERN-EP-2022-132",
    doi = "10.1007/JHEP06(2023)031",
    journal = "JHEP",
    volume = "06",
    pages = "031",
    year = "2023"
}

@article{Cowan:2010js,
    author = "Cowan, Glen and Cranmer, Kyle and Gross, Eilam and Vitells, Ofer",
    title = "{Asymptotic formulae for likelihood-based tests of new physics}",
    eprint = "1007.1727",
    archivePrefix = "arXiv",
    primaryClass = "physics.data-an",
    doi = "10.1140/epjc/s10052-011-1554-0",
    journal = "Eur. Phys. J. C",
    volume = "71",
    pages = "1554",
    year = "2011",
    note = "[Erratum: Eur.Phys.J.C 73, 2501 (2013)]"
}

\end{document}